\documentclass[letterpaper,aps,prd,twocolumn,tightenlines,preprintnumbers,nofootinbib,showkeys,superscriptaddress,longbibliography
]{revtex4-1}
\pdfoutput=1

\usepackage{XCharter}
\usepackage{mathptmx}
\usepackage{arydshln}
\usepackage[T1]{fontenc}
\usepackage{blkarray}
\usepackage{booktabs}
\usepackage{array}
\usepackage{fullpage}
\usepackage{amsfonts}
\usepackage{amsmath}
\usepackage{slashed}
\usepackage{amssymb}
\usepackage{graphicx}
\usepackage{epic}
\usepackage{eepic}
\usepackage{epsfig}
\usepackage{latexsym}
\usepackage[dvipsnames]{xcolor}
\usepackage[export]{adjustbox}
\usepackage{float}
\usepackage{multirow}
\usepackage{hyperref}
\usepackage{enumitem}
\usepackage{lipsum}
\hypersetup{colorlinks=true,citecolor=red,linkcolor=NavyBlue,urlcolor=NavyBlue}
\usepackage[caption=false,labelformat=simple]{subfig}
 
\usepackage{natbib}
\usepackage{relsize}
\usepackage[left=1.6cm,right=1.6cm,top=2.0cm,bottom=2.0cm]{geometry}
\usepackage{comment}
\usepackage{shorthand}
\usepackage{dcolumn}
\usepackage{longtable}
\usepackage{siunitx}

\usepackage{pifont}
\begin{document}


\title{Effects of a Brueckner-Hartree-Fock-corrected effective mass on speed of sound, conformality, and observables of dark matter-admixed neutron stars}

\author{Arijit Das}
\email{arijit21@iisertvm.ac.in}
\affiliation{Indian Institute of Science Education and Research Thiruvananthapuram, Vithura, Kerala, 695 551, India}

\author{Prashanth Jaikumar}
\email{prashanth.jaikumar@csulb.edu}
\affiliation{California State University Long Beach, Long Beach, California USA 90840}

\author{Adarsh Karekkat}
\email{adarsh.karekkat@unicaen.fr}
\affiliation{Université de Caen Normandie, ENSICAEN, CNRS/IN2P3, LPC Caen UMR6534, F-14000 Caen, France}
\affiliation{Indian Institute of Science Education and Research Thiruvananthapuram, Vithura, Kerala, 695 551, India}

\author{Tanumoy Mandal}
\email{tanumoy@iisertvm.ac.in}
\affiliation{Indian Institute of Science Education and Research Thiruvananthapuram, Vithura, Kerala, 695 551, India}

\begin{abstract}
\noindent
We construct an equation of state describing cold and dense matter in the core of neutron stars which includes an admixture of fermionic dark matter and incorporates nucleon effective masses derived from the relativistic Brueckner-Hartree-Fock (BHF) many-body approach within a relativistic mean-field model. Such a BHF-informed mixed-model approach increases stellar compactness, with mass-radius configurations which are consistent with smaller, lighter pulsars. The model displays the expected non-monotonic behaviour of sound speed hinted at by neutron-star data and is closer to the conformal bound at maximum mass. We find that the model displays tension with bounds on heavier pulsars, suggesting that the hypothesis of an aggregated dark component in neutron stars needs further critical study. 
\end{abstract}
\maketitle 


\section{Introduction}

Neutron stars (NSs) serve as natural astrophysical laboratories for studying strongly interacting matter at supernuclear densities.~Observational data—such as that gathered from gravitational wave (GW) events, pulsars, and other ultracompact objects (refer to Table~\ref{tab:Data} for a list of such experimental data)—help constrain the NS equation of state (EoS). Neither perturbative QCD (pQCD)~\cite{Somasundaram:2022ztm} nor lattice QCD~\cite{Philipsen:2012nu} can be reliably applied to describe strongly interacting matter at the intermediate densities found in the inner core of a NS. In this density regime, a useful approach is to extrapolate relativistic mean-field (RMF) models—fitted to empirical properties of finite nuclei and nuclear matter at saturation—with subsequent refinements, such as cross-couplings, density-dependent couplings, short-range correlations, or constraints from chiral effective-field theory (ChEFT)~\cite{Drischler:2021kxf}. 

\begin{table*}[hbt]
\centering
\renewcommand{\arraystretch}{1.1}
\setlength{\tabcolsep}{6pt}
\begin{tabular*}{\textwidth}{@{\extracolsep{\fill}}l c c c c@{}}
\toprule
\toprule
\textbf{Stellar object} & \textbf{$M/M_\odot$} & \textbf{$R$ (km)} & \textbf{$\Lambda$} & \textbf{$I$ ($10^{45} \, \mathrm{gm\, cm^2}
$)} \\
\midrule
\midrule
PSR J0030+0451 & $1.34^{+0.15}_{-0.16}$~\cite{Riley:2019yda,Miller:2019cac} & $12.71^{+1.14}_{-1.19}$~\cite{Riley:2019yda,Miller:2019cac} & $370^{+360}_{-130}$~\cite{Jiang:2019rcw} & $1.43^{+0.30}_{-0.13}$~\cite{Jiang:2019rcw} \\
PSR J0348+0432 & 1.97-2.05~\cite{Antoniadis:2013pzd} & 12.16-12.948~\cite{Huo:2018dqu} & $-$ & 1.594-1.9073~\cite{Zhao:2016rfv} \\
PSR J0740+6620 & $2.08\pm 0.07$~\cite{Fonseca:2021wxt} & $12.39^{+1.30}_{-0.98}$~\cite{Riley:2021pdl} & $-$ & $4.65^{+1.16}_{-0.82}$~\cite{Li:2022ozz} \\
PSR J1614-2230 & $1.97\pm 0.04$~\cite{Demorest:2010bx} & $8.6-13.8$~\cite{Majid:2020hlg} & $-$ & $-$ \\
PSR J0437–4715 & $1.418\pm 0.037$~\cite{Choudhury:2024xbk} & $11.36^{+0.95}_{-0.63}$~\cite{Choudhury:2024xbk} & $-$ & $-$ \\
PSR J1231-1411 & $1.12\pm 0.07$~\cite{Qi:2025mpn} & $9.91^{+0.88}_{-0.86}$~\cite{Qi:2025mpn} & $-$ & $-$ \\
PSR J0737-3039 A & $1.377\pm 0.005$~\cite{Lyne2006} & $8.14-25.74$~\cite{Yang} & $-$ & $1.15^{+0.38}_{-0.24}$~\cite{Landry:2018jyg} \\
PSR J0737-3039 B & $1.250\pm 0.005$~\cite{Lyne2006} & $-$ & $-$ & $-$ \\
GW170817 & 1.17-1.60~\cite{LIGOScientific:2017vwq} & $11.9\pm 1.4$~\cite{LIGOScientific:2018cki} & $70-580$~\cite{LIGOScientific:2017vwq} & $1.43^{+0.30}_{-0.13}$~\cite{Jiang:2019rcw} \\
GW190425 & $1.12-2.52$~\cite{LIGOScientific:2020aai} & $-$ & $1.4^{+3.8}_{-1.2}\times 10^3$~\cite{Han:2020qmn} & $-$ \\
\bottomrule
\bottomrule
\end{tabular*}
\caption{Summary of masses ($M/M_\odot$), radii ($R$), tidal deformabilities ($\Lambda$), and inferred moments of inertia ($I$) of various pulsars, GW events and other compact objects. The $M$-$R$ constraints for 4U 1724-207, 4U 1608-52 and EXO 1745-248 are taken from Ref.~\cite{Ozel:2015fia}. Similar constraints for HESS J1731-347 and PSR J1824-2452I (M28I) are taken from Refs.~\cite{Doroshenko:2022nwp} and~\cite{Vurgun:2022vvo} respectively. 
In our comparison with theoretical predictions, we have projected the moment of inertia (MoI) of PSR J0030+0451~\cite{Jiang:2019rcw}, estimated at a mass of $M = 1.4~M_\odot$, across the full range of mass and radius estimates~\cite{Riley:2019yda,Miller:2019cac}.}
\label{tab:Data}
\end{table*}

\par Studies within the frameworks of density-dependent relativistic mean-field theory~\cite{Hofmann:2000mc}, density-dependent relativistic Hartree-Fock theory~\cite{Sun:2008zzk}, and density-dependent quark mean-field theory~\cite{Huang:2024yjx} have demonstrated that the density dependence of the nucleon's effective mass~\cite{Li:2018lpy} impacts both the static and dynamic properties of NS. Hence, it is important to account for density-dependent nucleon masses when making precise predictions of NS observables. Within the RMF framework, such density-dependent mass ratios are typically obtained by self-consistently solving the equations of motion (EoM) of the underlying model. However, in the poorly understood intermediate baryon density range of $2\rho_0$--$5 \rho_0$ (where $\rho_0$ is the nuclear saturation density), many-body forces become increasingly important and may lead to structural changes in NS matter~\cite{Kojo:2020krb}. These effects have been shown to influence core-collapse supernovae and binary NS mergers~\cite{Li:2024tpr}, the cooling of NSs~\cite{Potekhin:2017ufy,Potekhin:2019eya}, and neutrino emissivity from NSs via the Urca process~\cite{Baldo:2014yja,DehghanNiri:2016cqm,Shternin:2018dcn}.

Given the pervasive importance of many-body effects, it is instructive to explore different ways of incorporating them into the RMF framework, albeit with a partial relaxation of the requirement for a fully self-consistent solution of the mean-field EoM. One way to implement this idea is to replace the (Dirac) effective nucleon mass in the RMF approach with an effective mass ratio obtained from realistic microscopic considerations. This represents an attempt to reconcile the RMF framework with more realistic in-medium nucleon interactions. For cold and dense matter, we adopt values of the nucleon effective mass calculated within the relativistic Brueckner-Hartree-Fock (BHF) framework~\cite{Sammarruca:2010hm}, which includes microscopic two-body and three-body forces~\cite{Zuo:2002sfa,Zuo:2002sg}. These results have also been used in studies of mass-radius relationships for purely hadronic NSs~\cite{Steiner:2011ft,Gandolfi:2013baa,Yamamoto:2017wre}. Moreover, as noted in Ref.~\cite{Reed:2025ccn}, a key limitation of many RMF models is their inability to support $\approx 2 M_\odot$ NSs, mainly due to the absence of three-body interactions in standard RMF frameworks. Introducing realistic, density-dependent meson-nucleon couplings that capture the effects of these three-body forces has been proposed as a solution to this issue. In a similar manner, a valid motivation for studying the impact of incorporating BHF-informed, density-dependent nucleon effective masses into RMF frameworks is to investigate their effect on the maximum mass of NSs.

\par Many-body interactions can leave their imprints on the speed of sound ($C_S$) profile, which, in turn, influences the stiffness of the EoS and thereby influences the maximum mass and radius of NS. The sound speed also provides insight into the microscopic composition of matter at supernuclear densities. Unfortunately, the limitations of theoretical approaches based on first principles result in large uncertainties in the sound speed profile at intermediate densities. At very low baryon densities, $C_S^2 \ll 1$~\cite{Tews:2018kmu}, while at asymptotically high densities, pQCD calculations predict a gradual approach to the conformal limit, $C_S^2 \rightarrow 1/3$. Moreover, studies have shown that, contrary to the holographic hypothesis of $C_S^2 \leq 1/3$~\cite{Cherman:2009tw} at intermediate densities, this value is not a strict upper bound, and the profile may become non-monotonic~\cite{Kojo:2020krb,Ecker:2017fyh,Altiparmak:2022bke,Brandes:2022nxa,Roy:2022nwy}. The holographic bound is in tension with observations of the most massive NS ~\cite{Bedaque:2014sqa}, as well as with predictions from quarkyonic matter~\cite{McLerran:2018hbz}, high-density QCD models~\cite{Leonhardt:2019fua}, and frameworks based on gauge-gravity duality~\cite{Kovensky:2021kzl}. The constraint $C_S^2 \leq 1/3$ significantly reduces the predicted NS mass below $2M_\odot$ for a class of non-relativistic and relativistic models, rendering them inconsistent with astrophysical observations~\cite{Moustakidis:2016sab}. Furthermore, in Ref.~\cite{Margaritis:2020onf} it has been shown that imposing the upper bound $C_S^2 = 1/3$ significantly decreases the maximum mass, Kerr parameter, and MOI of NSs, preventing these quantities from reaching values derived using realistic EoSs or constrained by observations. Relativistic kinetic theory also provides additional density-dependent constraints on the speed of sound~\cite{Olson:2000vx}. Given these uncertainties and constraints, it is instructive to ask whether the inclusion of many-body interactions in RMF models can shed further light on the behaviour of $C_S$.

\par An additional motivation for pursuing a more microscopic approach lies in the study of conformality in NS matter. Threshold values of parameters such as the polytropic index~\cite{Annala:2019puf}, conformal distance~\cite{Annala:2023cwx}, trace anomaly~\cite{Fujimoto:2022ohj}, and the curvature of the energy per particle~\cite{Marczenko:2023txe}, which characterize the onset of conformality, have already been determined. In contrast with model-agnostic frameworks—which emphasize using state-of-the-art theoretical and multimessenger constraints to provide confidence intervals for these parameters~\cite{Marczenko:2023txe}—we adopt a more microscopic approach by employing a BHF-informed model to study the emergence of conformality in NS.

\par We also go beyond conventional NSs by examining the accumulation of dark matter (DM) in their cores. Large amounts of accreted DM may produce observable effects on NS properties. In Ref.~\cite{Scordino:2024ehe}, a study of dark matter-admixed neutron stars (DMANS) was conducted within the BHF framework, starting from two-body and three-body nuclear interactions derived from ChEFT. However, since ChEFT is applicable only within a limited density range ($1\rho_0$--$2\rho_0$)~\cite{Tews:2018kmu}, it is reasonable to adopt a RMF model, informed by microscopic BHF results, to study the resulting properties of DMANS.

\par This paper is organized as follows. In Sec.~\ref{model}, we describe the model under consideration, present parameter sets and the EoS for hadronic and DMANS, compare the effective-mass ratios obtained from the RMF and BHF-informed models and discuss the crust-matching technique used to eliminate problematic regions of negative pressure in the EoS. Sec.~\ref{Result} presents numerical results for static and slowly rotating NSs and compares the predictions of the RMF and BHF-informed models for both hadronic NSs and DMANS. These results are also compared with experimental observations. In Sec.~\ref{result:conformality}, we examine the speed of sound profile and the relative degrees of conformality in the RMF and BHF-informed scenarios. Finally, Sec.~\ref{END} provides a summary and conclusions.

\section{Theoretical framework}
\label{model}

Among the many models proposed to describe NSs, we adopt the DM admixed $\mathrm{SU}(2)$ chiral sigma model~\cite{Sahu:1993db,Sahu:2000ut,Walecka:1974qa} for simplicity of illustration. Following Ref.~\cite{Guha:2021njn}, we present the relevant details of the model below.

\subsection{Model Lagrangian}

\noindent
The hadronic part of the model is given as follows~\cite{Sahu:2000ut}:
\begin{align}
    \mathcal{L}_{H} &= \overline{\psi}\slashed{D}\psi + \frac{1}{2}\bigg(\partial_{\mu}\vec{\pi}\cdot\partial^{\mu}\vec{\pi} + \partial_{\mu}\sigma\partial^{\mu}\sigma\bigg) - \frac{1}{4}\Omega_{\mu\nu}\Omega^{\mu\nu} \nn \\
    &+ \frac{1}{2}g_{\omega}^2x^2 \omega_{\mu}\omega^{\mu} - \frac{1}{4}\vec{R}_{\mu\nu}\cdot\vec{R}^{\mu\nu} + \frac{1}{2}m^2_{\rho}\vec{\rho}_{\mu}\cdot\vec{\rho}^{\mu} -U(\bar{x}).
\end{align}
Here, $\psi$ is the $\mathrm{SU}(2)$ nucleon isospin doublet. The operator $\slashed{D}$ is defined as,
\begin{align}
    \slashed{D} &= i\gamma_{\mu}\partial^{\mu} - g_{\sigma}(\sigma + i\gamma_{5}\vec{\tau}\cdot\vec{\pi})
    - g_{\omega}\gamma_{\mu}\omega^{\mu} 
    - \frac{1}{2}g_{\rho}\gamma_{\mu}\vec{\tau}\cdot\vec{\rho}^{\mu}, 
\end{align}
with $\pi$, $\sigma$, $\omega$ and $\rho$ are the mesonic fields, and $\Omega_{\mu\nu} = \partial_{\mu}\omega_{\nu} - \partial_{\nu}\omega_{\mu}$ and $\vec{R}_{\mu\nu} = \partial_{\mu}\vec{\rho}_{\nu} - \partial_{\nu}\vec{\rho}_{\mu} + g_{\rho}(\vec{\rho}_{\mu}\times \vec{\rho}_{\nu})$ are the field strength tensors for $\omega$ and $\rho$ mesons, respectively. The mass of $\rho$ is denoted by $m_\rho$, and $g_\sigma$, $g_\omega$ and $g_\rho$ are the coupling constants. The scalar potential $U(\bar{x})$ is defined as
\begin{align}
\label{eq:sclrpot}
    U(\bar{x}) = \frac{\lambda}{4}\bar{x}^{4} + \frac{B}{6}\bar{x}^{6} + \frac{C}{8}\bar{x}^{8},
\end{align}
where $\bar{x}^2 = x^2 - x^2_0$ with $x^2 = \sigma^2 + \vec{\pi}^2$, and $x_0$ being the vacuum expectation value (vev) of $\sigma$. Here, $\lambda$, $B$, and $C$ are some constants in the potential.

The fermionic dark sector Lagrangian is given by
\begin{align}
    \mathcal{L}_{DM} = \overline{\chi}\left( i\gamma_{\mu}\partial^{\mu} - m_{\chi} \right)\chi .
\end{align}
Here, $\chi$ is the DM spinor and $m_{\chi}$ is its mass.

The new scalar mediator $\phi$ and the new vector mediator $\xi$ connect the hadronic sector and the dark sector via the following interactions,
\begin{align}
    \mathcal{L}_{I} = - g_{\phi}\overline{\psi}\psi\phi - g_{\xi}\overline{\psi}\gamma_{\mu}\psi \xi^{\mu} -y_{\phi}\overline{\chi}\chi\phi -y_{\xi}\overline{\chi}\gamma_{\mu}\chi \xi^{\mu}.
\end{align}
Here, $g_{\phi}$, $g_{\xi}$, $y_{\phi}$ and $y_{\xi}$ are the interaction couplings. The kinetic and mass terms for the new mediators $\phi$ and $\xi$ are represented in the following Lagrangian
\begin{align}
    \mathcal{L}_{M} = -\frac{1}{4}\Xi_{\mu\nu}\Xi^{\mu\nu} + \frac{1}{2}m^{2}_{\xi}\xi_{\mu}\xi^{\mu} + \frac{1}{2}\partial_{\mu}\phi\partial^{\mu}\phi - \frac{1}{2}m^2_{\phi}\phi^2,
\end{align}
where the masses for $\xi$ and $\phi$ are denoted by $m_\xi$ and $m_\phi$ respectively and $\Xi_{\mu\nu} = \partial_{\mu}\xi_{\nu} - \partial_{\nu}\xi_{\mu}$ is the field strength tensor for $\xi_{\mu}$.

The leptonic sector is completely decoupled from other sectors of the model and given by 
\begin{align}
    \mathcal{L}_{\ell} = \sum_{\ell}\overline{\ell}\left( i\gamma_{\mu}\partial^{\mu} - m_{\ell} \right)\ell 
\end{align}
Here, $\ell$ ($=e,\mu$) is the spinor field for the charge lepton of flavour $\ell$ of mass $m_{\ell}$. 

Collecting all the terms, the total model is given by
\begin{align}
    &\mathcal{L}  = \mathcal{L}_{H} + \mathcal{L}_{DM} + \mathcal{L}_{I} + \mathcal{L}_{M} + \mathcal{L}_{\ell}.
\end{align}
   
The masses of the nucleons ($m_N$ where $N=p,n$), $\omega$ meson ($m_{\omega}$), and $\sigma$ meson ($m_{\sigma}$) arise from chiral symmetry breaking. If $\phi_0$ denotes the vev of the scalar field $\phi$, then the masses are given by
\begin{align}
m_N = g_{\sigma}x_{0} + g_{\phi}\phi_0,\quad
m_{\omega} = g_{\omega}x_{0},\quad
m_{\sigma} = \sqrt{2\lambda}x_0.
\end{align}
Due to the presence of the scalar mediator $\phi$, the nucleon mass now receives an additional contribution from $\phi_0$.

\subsection{Equation of state}

In the RMF approximation, one replaces the meson fields by their RMF expectation values and, hence, all the corresponding kinetic terms drop out of the RMF approximated model. The effective masses of nucleons and the DM are then given by
\begin{align}
m_N^{\ast} = g_{\sigma}\sigma + g_{\phi}\phi,\quad
m^{*}_{\chi} = m_{\chi} + y_{\phi}\phi,
\end{align}
where in the above two expressions, $\phi$ and $\sigma$ are now RMF values of those fields. In all the following discussions, except that for fermions, all other fields will be assumed to have their RMF values. Before presenting the EoS, we provide the reader with some necessary relations. The scalar density for a generic fermion ($f=N,\chi,\ell$) is given by
\begin{align}
    \rho^{S}_f = \langle \overline{f}f \rangle &= \frac{\gamma_f}{2\pi^2}\int_{0}^{k_{f}}dk \frac{m_f^{\ast}k^2}{\sqrt{k^2 + m_f^{\ast 2}}} \nn \\
    &= \frac{\gamma_f m_f^{\ast 3}}{4\pi^2}\left[x_{f}\sqrt{1 + x^{2}_{f}} - \sinh^{-1}x_{f} \right].
\end{align}
\noindent
Here, $\gamma_f = 2$ is the spin degeneracy factor and $x_{f} = k_{f}/m_f^{\ast}$ where $k_f$ is the Fermi momentum. The vector densities \cite{Glendenning:1997wn} are given by
\begin{align}
\label{rho_v}
    \rho_f^V = \langle f^{\dagger}f \rangle = \langle \overline{f}\gamma^{0}f \rangle = \frac{\gamma_{f}}{2\pi^2}\int_{0}^{k_{f}}dk~k^2 = \frac{\gamma_{f}}{6\pi^2}k_{f}^3\,.
\end{align}
The general expressions (for a generic fermion $f=n, p, \ell, \chi$) for the pressure ($\mathcal{P}$) and energy density ($\mathcal{E}$) can be obtained as follows
\begin{align}
\mathcal{P}_f &= \frac{\gamma_f}{6\pi^2}\int_{0}^{k_{f}}dk\frac{k^4}{\sqrt{k^2 + m_f^{\ast 2}}} \nn \\
&= \frac{\gamma_f m_f^{\ast 4}}{48\pi^2}\left[(-3x_{f} + 2x^3_{f})\sqrt{1 + x^2_{f}} + 3\sinh^{-1}x_{f} \right], \\ \nn \\
\mathcal{E}_f
&= \frac{\gamma_f}{2\pi^2}\int_{0}^{k_{f}}dk~
     k^2 \sqrt{k^2 + m_f^{\ast 2}} \nn \\
&= \frac{\gamma_f m_f^{\ast 4}}{16\pi^2}\left[(x_{f} + 2x^3_{f})\sqrt{1 + x^2_{f}} - \sinh^{-1}x_{f} \right].
\end{align}

\begin{table*}[!htbp]
\centering
\renewcommand{\arraystretch}{1.5}
\footnotesize
\setlength{\tabcolsep}{4pt} 
\resizebox{\textwidth}{!}{
\begin{tabular}{llccccccccccccc}
\toprule
\toprule
\multicolumn{15}{c}{\textbf{Hadronic sector}} \\
\midrule
\midrule
& Parameter & $C_{\sigma}$ & $C_{\omega}$ & $C_{\rho}$ & $B$ & $C$ & $m_{\sigma}$ & $f_{\pi} = x_0$ & $K$ & $B.E./A$ & $J$ & $L_0$ & $\rho_0$ & $Y(\rho_0)$ \\
& Units      & (fm$^2$)     & (fm$^2$)     & (fm$^2$)   & (fm$^2$) & (fm$^4$) & (MeV) & (MeV) & (MeV) & (MeV) & (MeV) & (MeV) & (fm$^{-3}$) & -- \\
& Value     & 7.325        & 1.642        & 5.324      & $-$6.586 & 0.571    & 444.614 & 153.984 & 231 & $-$16.3 & 32 & 88 & 0.153 & 0.87 \\
\bottomrule
\bottomrule
\end{tabular}%
}
\caption{Parameter set for hadronic sector \cite{Guha:2021njn,Jha:2008yth}.}
\label{tab:had_par}
\end{table*}

\vspace{0.5em}

\begin{table*}[!htbp]
\centering
\renewcommand{\arraystretch}{1.4}
\scriptsize
\setlength{\tabcolsep}{2.5pt} 
\resizebox{0.52\textwidth}{!}{
\begin{tabular}{llccccccccc}
\toprule
\toprule
\multicolumn{11}{c}{\textbf{Dark matter sector}} \\
\midrule
\midrule
& Parameter & \multicolumn{2}{c}{$m_{\chi}$ (MeV)} &  \multicolumn{2}{c}{$m_{\phi}$ (MeV)} & \multicolumn{2}{c}{$m_{\xi}$ (MeV)} & \multicolumn{2}{c}{$y_{\phi} = y_{\xi}$} & $g_{\phi} = g_{\xi}$ \\
& BP I   & \multicolumn{2}{c}{5}   & \multicolumn{2}{c}{9}    & \multicolumn{2}{c}{11}   & \multicolumn{2}{c}{0.13} & $1.1 \times 10^{-4}$ \\
& BP II  & \multicolumn{2}{c}{15}  & \multicolumn{2}{c}{20}   & \multicolumn{2}{c}{34}   & \multicolumn{2}{c}{0.21} & $1.1 \times 10^{-4}$ \\
\bottomrule
\bottomrule
\end{tabular}%
}
\caption{Parameter sets for DM sector \cite{Guha:2021njn,Sen:2021wev}.}
\label{tab:dark_par}
\end{table*}

\vspace{0.5em}

\begin{table*}[!htbp]
\centering
\renewcommand{\arraystretch}{1.4}
\scriptsize
\setlength{\tabcolsep}{2.5pt} 
\resizebox{0.52\textwidth}{!}{
\begin{tabular}{llcccccccc}
\toprule
\toprule
\multicolumn{10}{c}{\textbf{BHF Fitting Parameters}} \\
\midrule
\midrule
& Parameter & $a_1$ & $b_1$ & $c_1$ & $a_2$ & $b_2$ & $c_2$ & $d_1$ & $d_4$ \\
& Value     & 0.102 & $-$0.094 & 0.068 & 0.699 & 0.0354 & 0.0133 & $-$0.0135 & $-$0.367 \\
\bottomrule
\bottomrule
\end{tabular}%
}
\caption{BHF fitting parameters \cite{Shang:2020kfc}.}
\label{tab:bhf_par}
\end{table*}

\begin{figure*}
\centering
\captionsetup[subfigure]{labelformat=empty}
\subfloat[\quad(a)]{\includegraphics[width=0.5\textwidth]{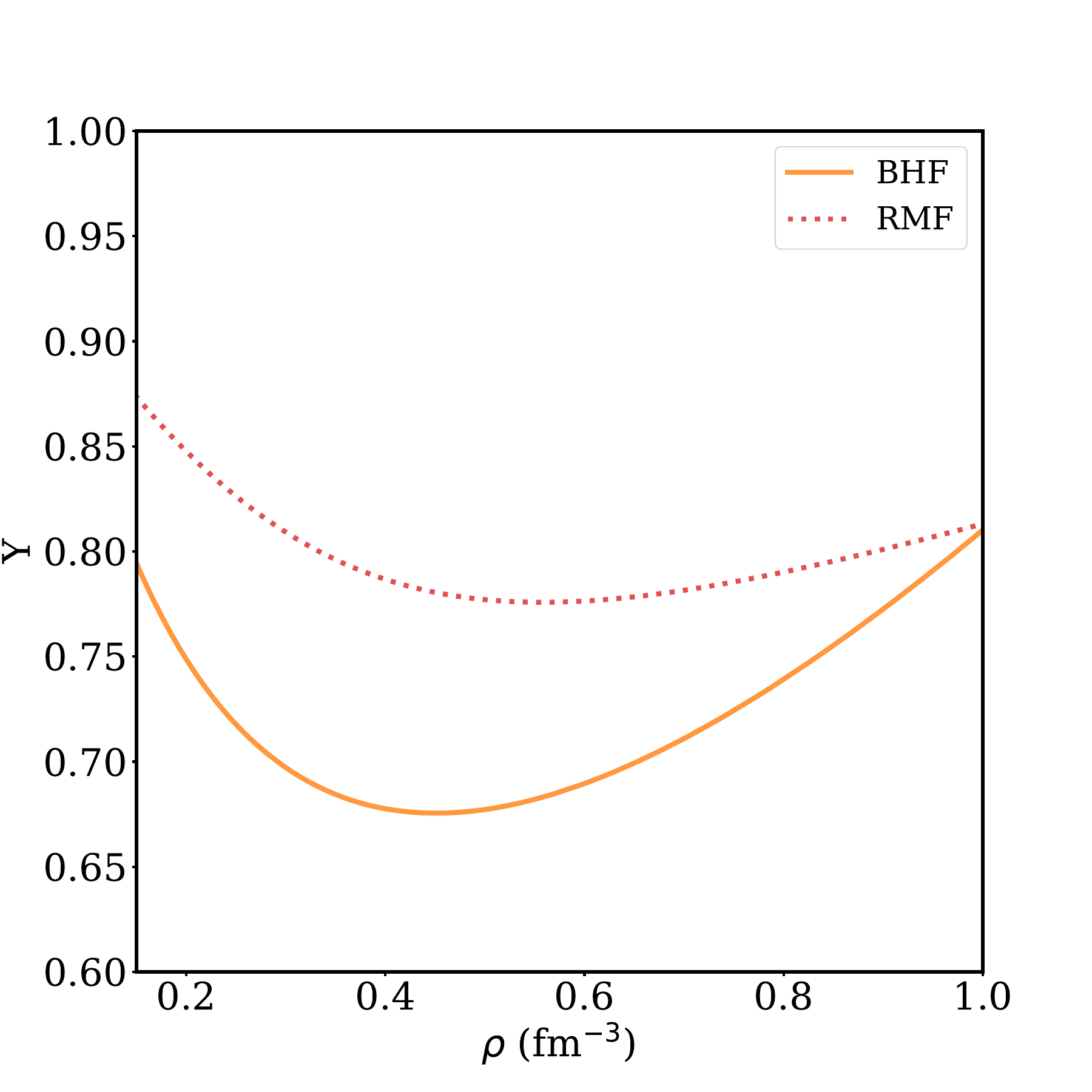}\label{BHF_Y}}
\subfloat[\quad(b)]{\includegraphics[width=0.5\textwidth]{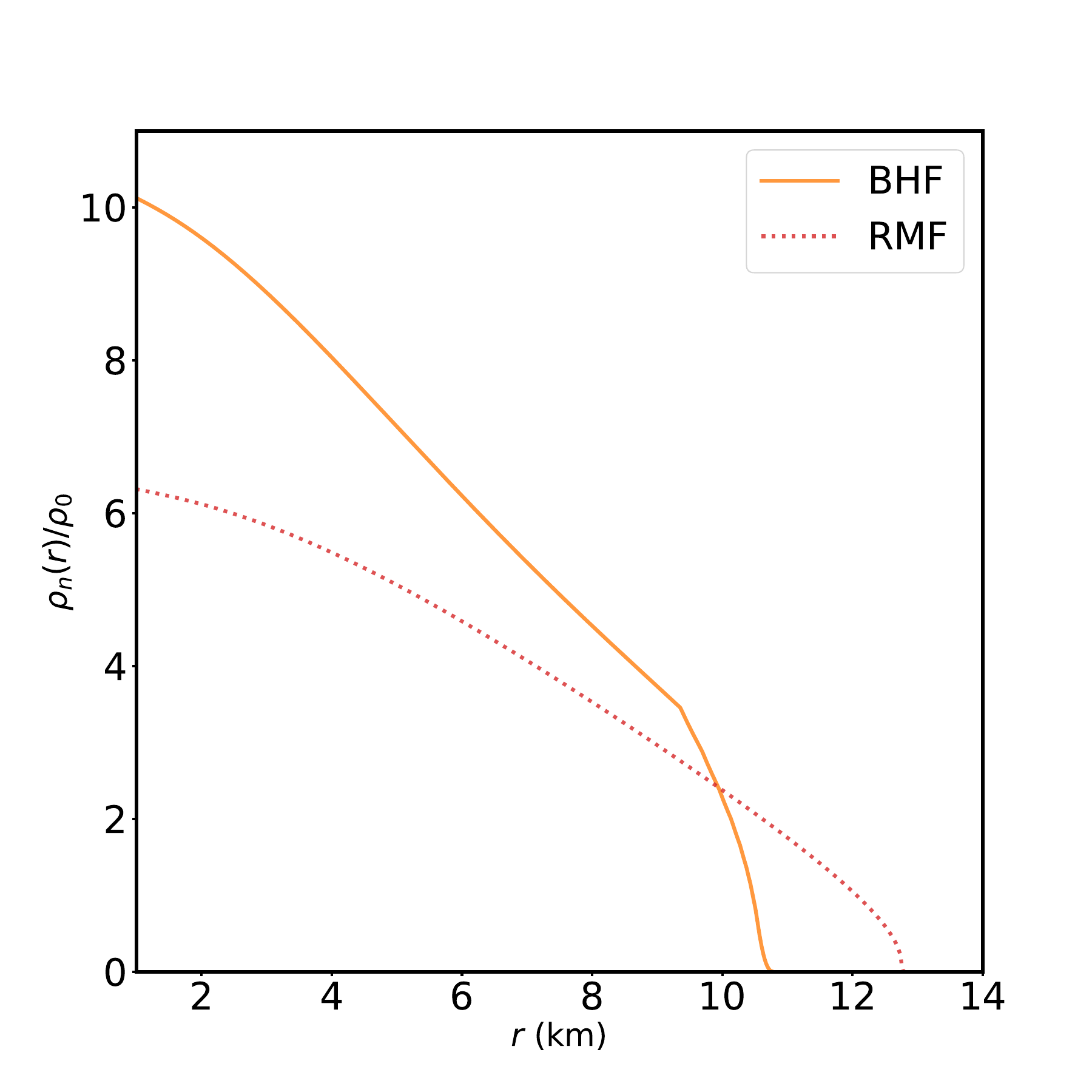}\label{fig:n_rad_var}}
\caption{(a) Comparison of RMF and BHF corrected nucleon effective mass ratios for purely hadronic EoS. (b) The density profiles $\rho(r)/\rho_0$  for RMF and BHF hadronic $2.0 M_\odot$ NS. The higher density in the BHF case leads to a more compact NS. \label{fig:RMF-BHF-comp}}
\end{figure*}

The expressions for the total pressure and energy density of DMANS matter are given as follows:
\begin{align}
\label{eq:PE}
\mathcal{P} = \sum_{f = N, \chi, \ell}\mathcal{P}_f -\Sigma + \Theta_{-},\quad
\mathcal{E} = \sum_{f = N, \chi, \ell}\mathcal{E}_f +\Sigma + \Theta_{+} 
\end{align}
The expressions for the quantities $\Sigma$ and $\Theta_{\pm}$ can be found from Eqs.~\eqref{Cap_sigma} and \eqref{Cap_Theta} in Appendix \ref{app:eos_eom}.
In the above expressions, the density-dependent RMF fields $\sigma(\rho^V)$ and $\phi(\rho^V)$ ($\rho^V = \rho^V_n + \rho^V_p$) obtained from self-consistent solutions of Eqs.~\eqref{eom:phi} and \eqref{eom:sigma} (refer to Appendix \ref{app:eos_eom} for discussions on the complete set of EoMs) should be used in order to numerically evaluate the EoS. 

\subsection{Parameter sets and simplified EoS}
\label{PSET}

There are five free parameters in the hadronic part of the model i.e. $C_{\sigma} = g^2_{\sigma}/m^2_{\sigma}$, $C_{\omega} = g^2_{\omega}/m^2_{\omega}$, $C_{\rho} = g^2_{\rho}/m^2_{\rho}$, $B$, and $C$. These parameters of the hadronic sector are determined by fitting to symmetric nuclear matter (SNM) properties as shown in Table~\ref{tab:had_par}~\cite{Guha:2021njn,Jha:2008yth}. The tabulated values of SNM properties e.g. saturation density ($\rho_0 = 0.153$~fm$^{-3}$), nucleon effective mass at satuartion [$Y(\rho_0) = 0.87$], nuclear incompressibility ($K\approx 231$ MeV), binding energy per nucleon ($E_{BE}/A = -16.3$~MeV) and symmetry energy coefficient ($J = 32$~MeV), calculated using the specified parameter set, agree well with already existing constraints. The value of the slope parameter at saturation ($L_0=88$) also falls well within already established constraints~\cite{Dutra:2014qga}. We use the parameter set for the DM sector presented in Table~\ref{tab:dark_par}. These values fall very well in agreement with the self-interaction cross-section constraints from the Bullet Cluster and the present thermal relic abundance of DM~\cite{Tulin:2013teo}. The couplings of the new mediators with the nucleons, $g_{\phi}$ and $g_{\xi}$, are much smaller than the corresponding couplings with the DM, $y_{\phi}$ and $y_{\xi}$. Therefore, the DM-nucleon scattering cross section agrees well with the available constraints~\cite{Randall:2008ppe}. In all our further considerations, we only consider NS composed solely of neutrons and DM [i.e. in Eq.~\ref{eq:PE}, $f$ will run over $n$ and $\chi$ only]. We also assume the following benchmark values to present our results: $M_{\chi} = 5$ and $15$~GeV and $k_{\chi}= 80$~MeV. We will call these DM sector parameter sets BPI and BPII respectively. The EoS $\mc{P} = \mc{P}(\mc{E})$ can be obtained by numerically eliminating $k_n$ from the expressions of $\mc{P}$ and $\mc{E}$.

\subsection{RMF vs BHF: effective masses}
\label{EFT_BHF_Comp}

As pointed out earlier, in the purely RMF-based approach to the DMANS model, Eqs.~\eqref{eom:phi} and \eqref{eom:sigma} must be solved self-consistently over a realistic range of $k_n$, with a fixed $k_{\chi}$, to obtain $\phi = \phi(\rho^{V})$ and $\sigma = \sigma(\rho^{V})$ as functions of baryon density. Consequently, the effective neutron mass $m^{\ast}_n$ and the DM effective mass $m^{\ast}_{\chi}$ both become functions of $\rho^{V}_{n}$. In all numerical calculations that follow in the RMF scenario, the functions $\phi(\rho^{V})$, $m^{\ast}_n(\rho^{V})$, and $m^{\ast}_{\chi}(\rho^{V})$ must be used. In contrast, in the BHF scenario, we adopt the following approximate analytical fitting function for the effective-mass ratio at zero temperature from Ref.~\cite{Shang:2020kfc}.
\begin{align} 
\label{eqn:YBHF}
Y_{\mathrm{BHF}} &= a_1 + b_1 \beta + c_1 \beta^2 + \left(a_2 + b_2 \beta + c_2 \beta^2\right)\rho^{V} \nn \\
&+ \dfrac{d_1}{\rho^{V}} + d_4\ln \rho^{V}.
\end{align}
The values of the fitting parameters $a_{i}$, $b_{i}$, $c_{i}$, and $d_{i}$ are presented in Table~\ref{tab:bhf_par}. The asymmetry parameter is $\beta = 1$ in our case, since it is defined as $\beta = \left(\rho^V_n - \rho^V_p\right)/\left(\rho^V_n + \rho^V_p\right)$. This fitting function is valid within the range $\rho_n^V \in [0.1, 0.8]~\text{fm}^{-3}$. In the BHF scenario, we use $m^{\ast}_n = Y_{\mathrm{BHF}} m_n $ as the modified nucleon effective mass in Eq.~\eqref{eom:phi}, and solve for $\phi = \phi_{\mathrm{BHF}}(\rho^V)$. Consequently, the BHF-corrected effective DM mass becomes $m^{\ast}_{\chi}(\rho^V) = m_{\chi} + y_{\phi}\phi_{\mathrm{BHF}}(\rho^V)$. In Fig.~\ref{BHF_Y}, we compare the mass ratio $Y$ computed from the RMF approach with $Y_{\mathrm{BHF}}$ for purely hadronic matter. One can clearly see that the BHF correction deviates significantly from the RMF mass ratio within the validity range of the BHF framework. Hence, notable differences in the EoS and macroscopic observables of NS are expected. In Fig.~\ref{fig:n_rad_var}, we show the radial density profiles $\rho(r)/\rho_0$ of 2$M_\odot$ NSs in both the RMF and BHF scenarios. These have been obtained by solving the Tolman-Oppenheimer-Volkoff (TOV) equations as shown in Eq.~\eqref{eq:tov} in Appendix~\ref{macro}. The density is noticeably higher in the BHF case, resulting in a smaller radius for NS of the same mass compared with the RMF case.

\par Finally, we emphasize the choice of the BHF effective mass used in the following numerical analyses. Both the pressure $\mc{P}$ and the energy density $\mc{E}$ are evaluated at various values of the neutron Fermi momentum $k_n$, i.e., on different Fermi surfaces. On each Fermi surface, the Landau effective mass provides a more accurate characterization of quasi-particle dynamics in the presence of interactions, encapsulating modifications arising from two- and three-body forces. Therefore, to better account for the influence of three-body interactions in our numerical analysis for the BHF case, we employ the Landau effective mass rather than the Dirac effective mass.
\par In evaluating the BHF-corrected EoS, the energy density $\mathcal{E}(r=R)$ at the stellar surface--defined by $\mathcal{P}(r=R)=0$--does not vanish. This may create an issue for the standard calculation of the tidal Love number $k_2$, which is required to determine the tidal deformability $\Lambda$. In such cases, a crust-matching technique can be employed at low energy densities and pressures to reproduce crustal behaviour using a reliable crust EoS \cite{Abac:2021txj}. Here, we adopt the well-known BSK19 crust EoS, matching it to the low-energy-density portion of our EoS. We locate the energy density at which the crust EoS intersects the hadronic (or DMANS) EoS--namely, the point where their pressures are equal--and then construct a piecewise-continuous EoS: to the left of the intersection, we replace the original EoS with the BSK19 crust branch, while to the right we retain the hadronic or DMANS branch, as appropriate.

\section{Observables and comparison with data}
\label{Result}

\begin{figure}[t]
    \centering
    \includegraphics[width=0.985\linewidth]{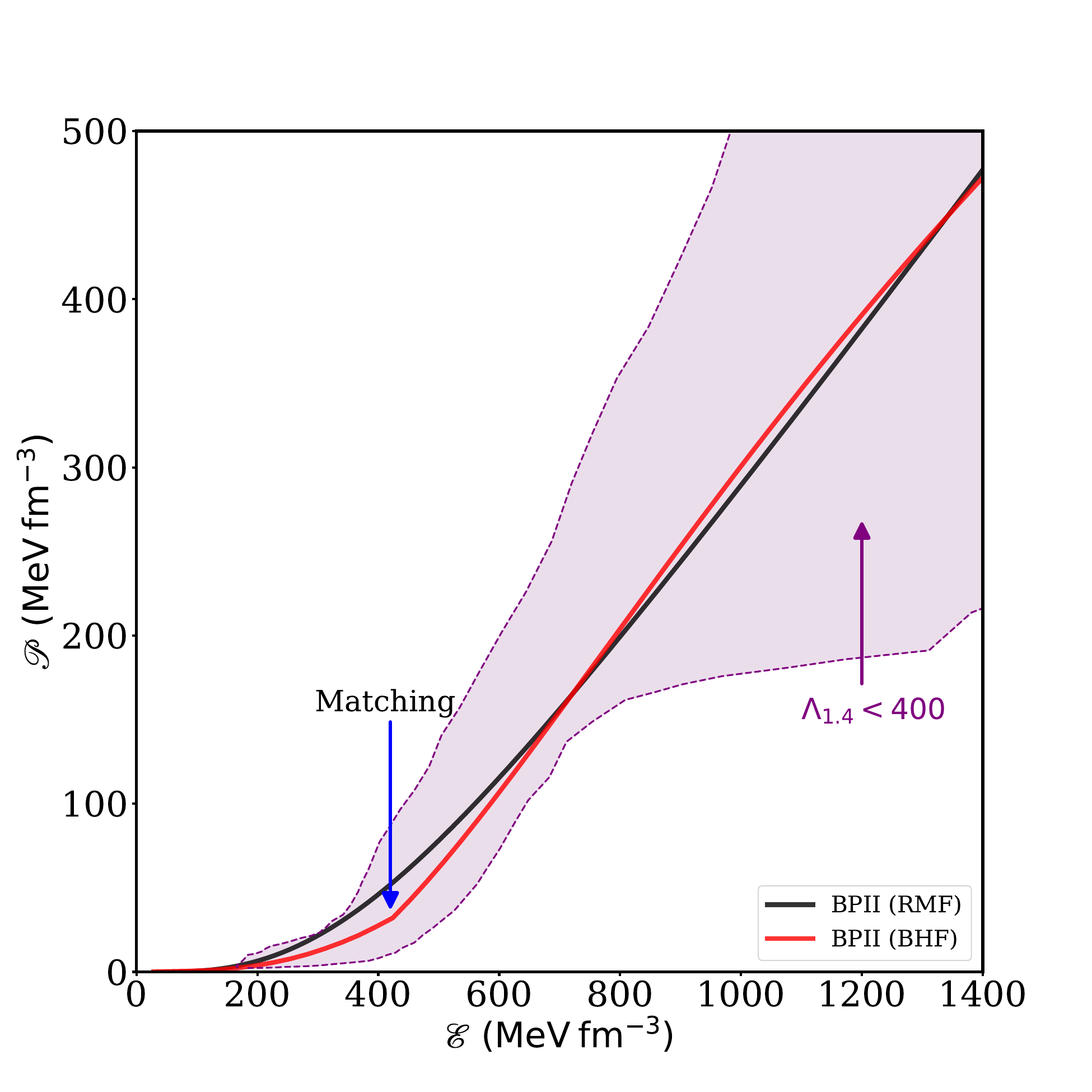}
    \caption{Comparison of EoSs for RMF and BHF corrected cases for BP II parameter set. The constraint $\Lambda_{1.4} < 400$ is taken from Ref.~\cite{Annala:2017llu}.}
    \label{fig:EOS_Comp_15_60}
\end{figure}

\begin{figure*}
\centering
\captionsetup[subfigure]{labelformat=empty}
\subfloat[\quad(a)]{\includegraphics[width=0.5\textwidth]{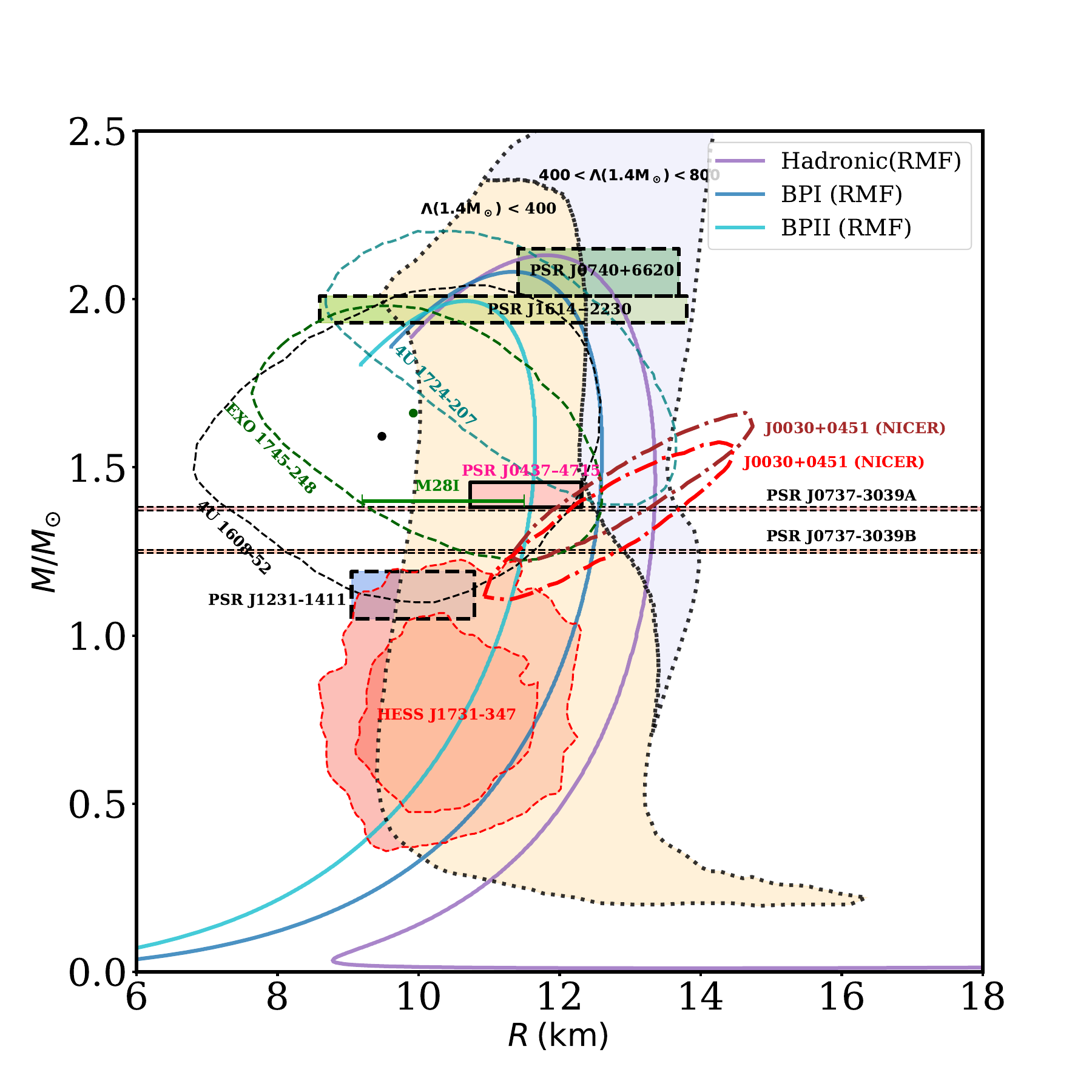}\label{rx-my-pulsar-rmf}}
\subfloat[\quad(b)]{\includegraphics[width=0.5\textwidth]{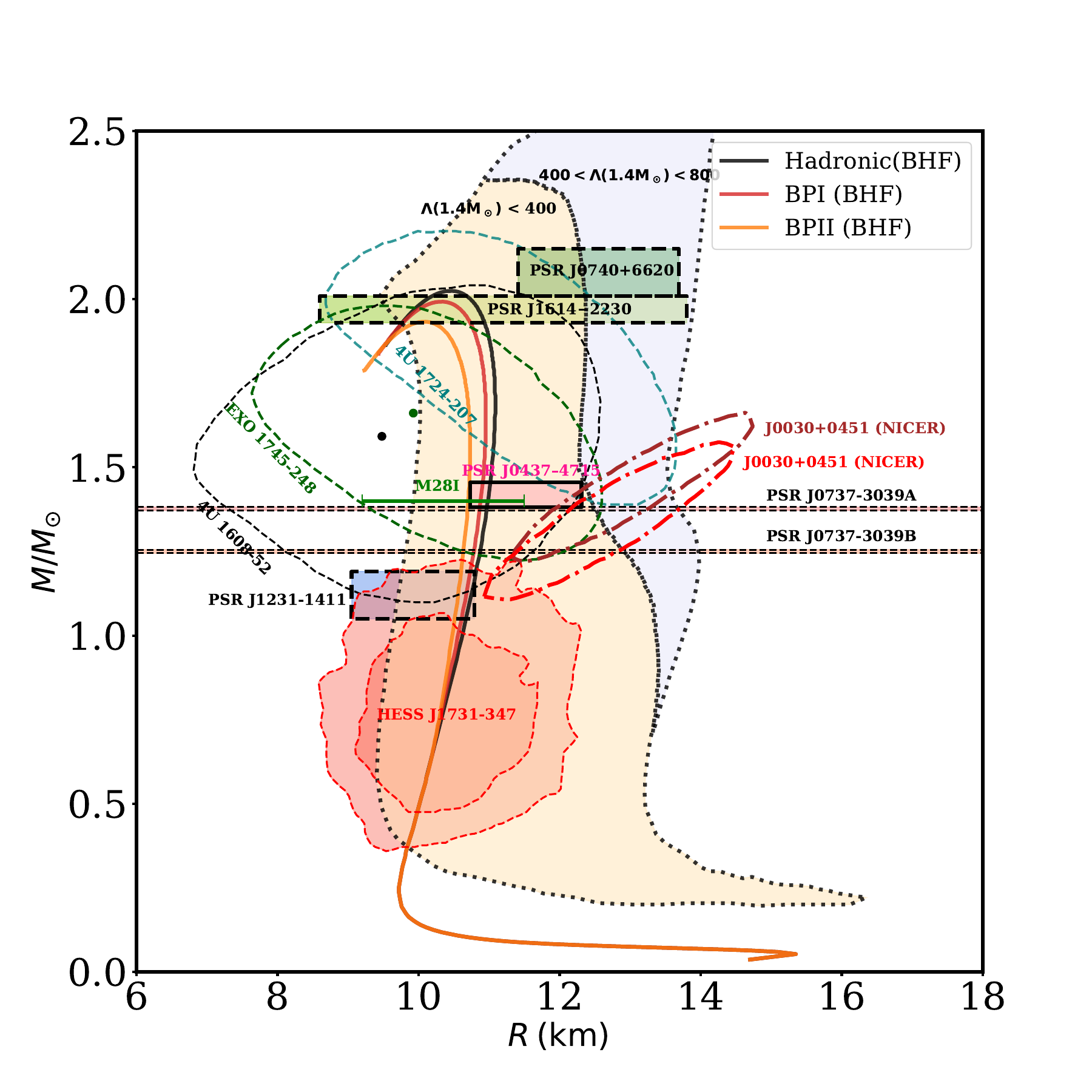}\label{rx-my-pulsar-bhf}}
\caption{(a) Mass-radius relationships for hadronic, BPI and BPII parameters in the RMF scenario. Theoretical results are compared with relevant experimental results presented for pulsars in Table \ref{tab:Data}. (b) Mass-radius relationships for hadronic, BPI and BPII parameters in the BHF scenario. The M-R curves in the BHF scenario are completely within the $\Lambda(1.4M_\odot) < 400$ constraint obtained from \cite{Annala:2017llu}, which is consistent with the result in Fig.\ref{fig:M-Lambda}.
\label{fig:rx_my}}
\end{figure*}

\begin{figure}[t]
    \centering
    \includegraphics[width=1.0\linewidth]{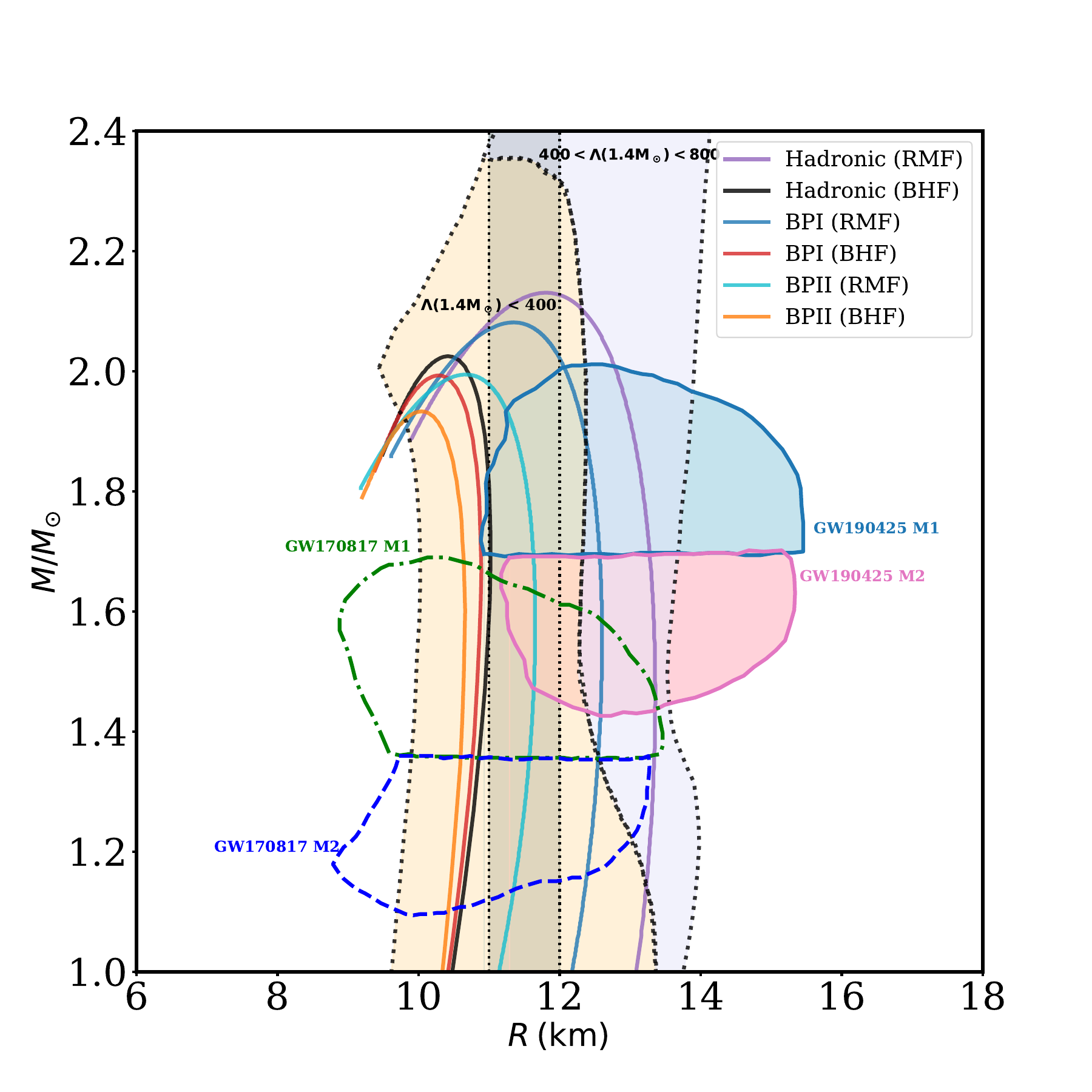}
    \caption{Mass-radius relationships for hadronic, BPI and BPII parameters in the both RMF and BHF scenarios are compared with various GW waves data.}
    \label{rx-vs-my-gw}
\end{figure}

\begin{figure*}
\centering
\captionsetup[subfigure]{labelformat=empty}
\subfloat[\quad(a)]{\includegraphics[height=9.0cm,width=9.0cm]{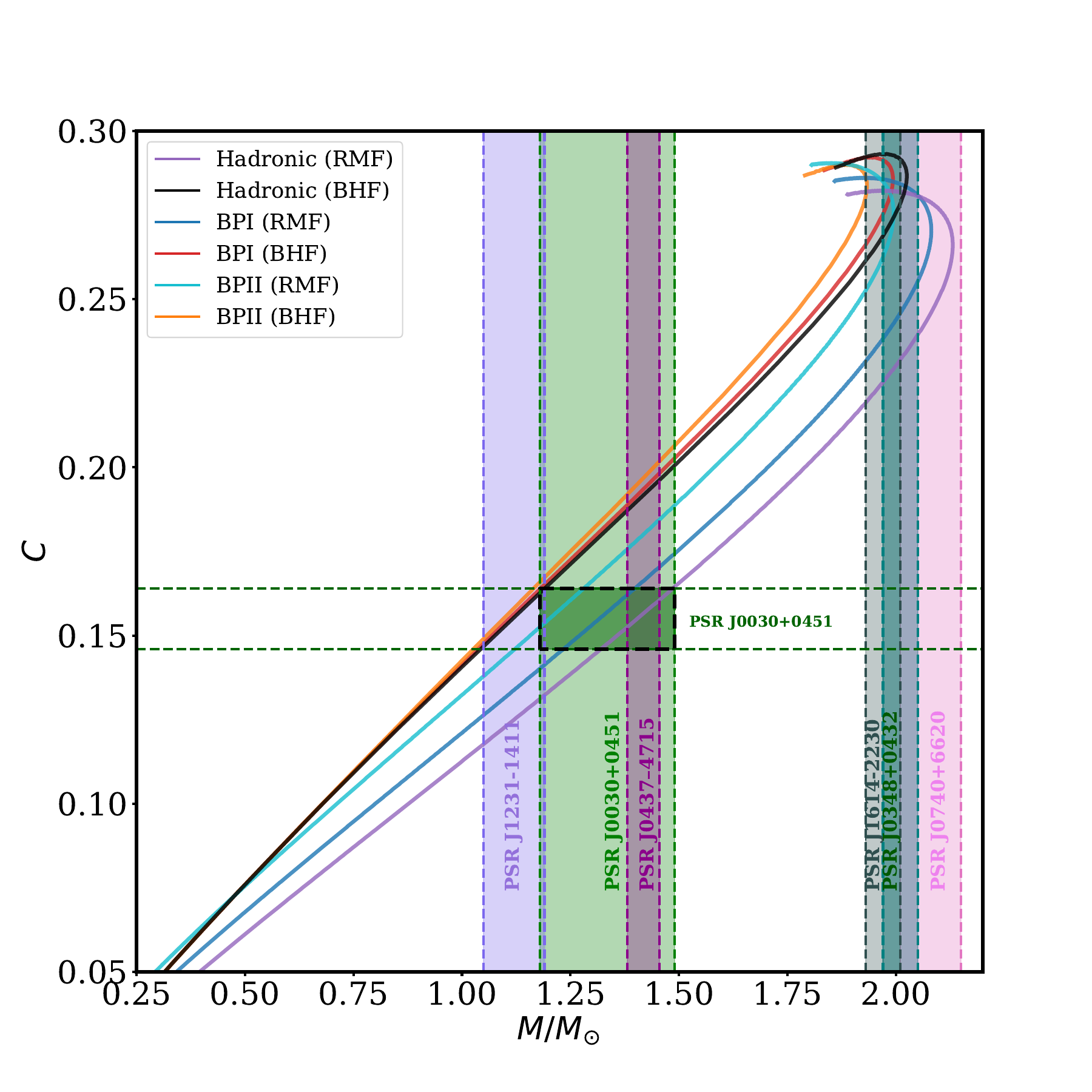}\label{fig:M-C}}\hfill
\subfloat[\quad(b)]{\includegraphics[height=9.0cm,width=9.0cm]{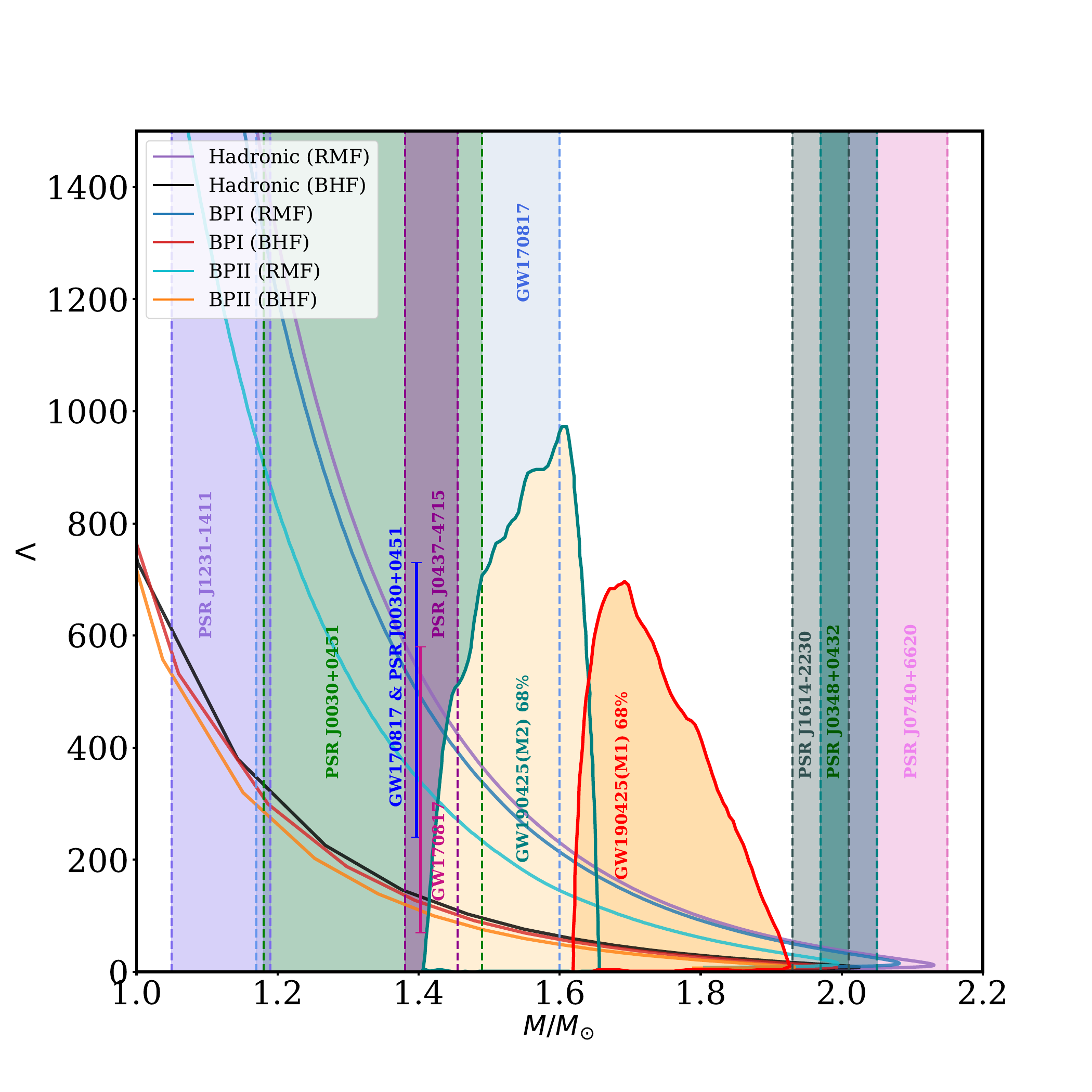}\label{fig:M-Lambda}}\\
\subfloat[\quad(c)]{\includegraphics[height=9.0cm,width=9.0cm]{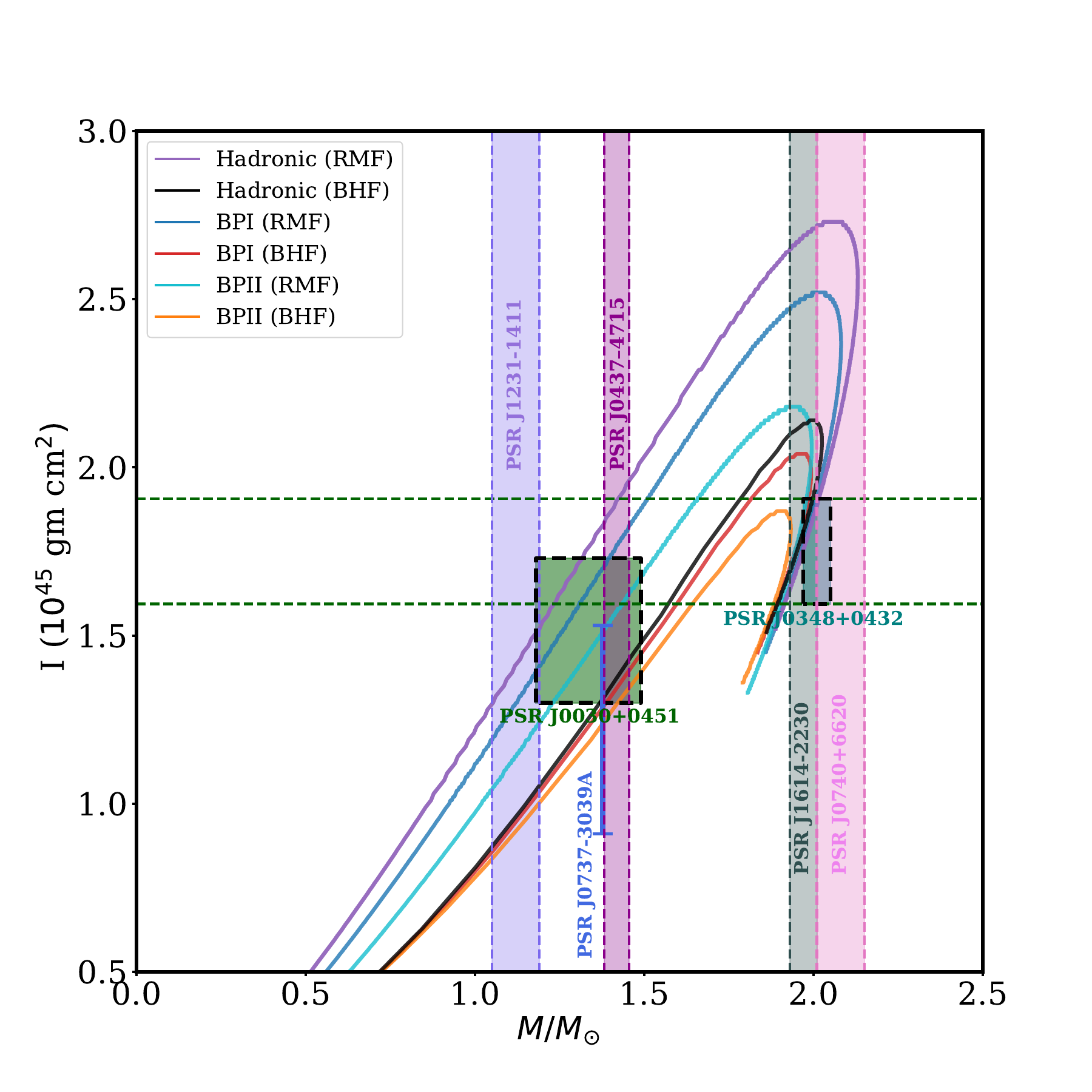}\label{fig:M_I(a)}}\hfill
\subfloat[\quad(d)]{\includegraphics[height=9.0cm,width=9.0cm]{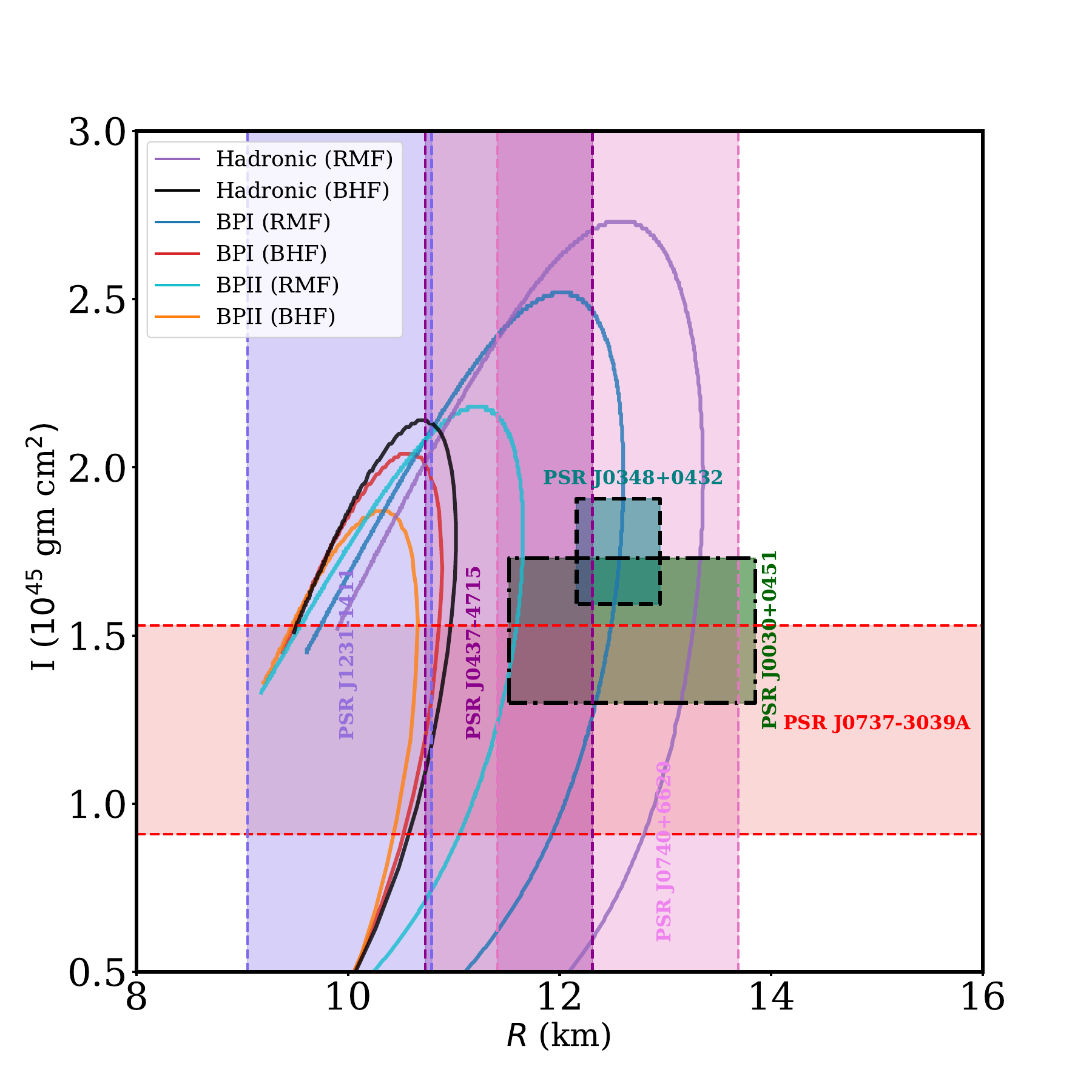}\label{fig:R_I(b)}}
\caption{Compactness ($C$), Tidal deformability ($\Lambda$) and MoI ($I$) of NSs as functions of mass ($M$ in $M_\odot$ units) are presented in panels (a), (b), and (c), respectively. Panel (d) shows the variation of $I$ with stellar radius ($R$). Results obtained for purely hadronic and DMANS (for BPI and BPII) cases in RMF and BHF scenarios are compared with the experimental findings presented in Table~\ref{tab:Data}.}
\label{fig:M_I}
\end{figure*}

\par In Fig.~\ref{fig:EOS_Comp_15_60}, we show a comparison of EoSs for the BPII parameter set obtained within the RMF and BHF approaches. The EoS lines are passing through a colored region which represents the constraint, $\Lambda_{1.4} < 400$ (where $\Lambda_{1.4}$ is the value of dimensionless tidal deformability $\Lambda$ at the benchmark mass of $M = 1.4M_\odot$), as obtained in Ref.~\cite{Annala:2017llu}. In Ref.~\cite{Annala:2017llu}, an ensemble of EoS are generated with (i) ChEFT EoS for $\rho < 1.1\rho_0$, (ii) piecewise polytropes between the regions of validity of ChEFT and pQCD and (iii) NNLO pQCD results in the high density regime to describe deconfined quark matter. To obtain this constraint, the condition of existence of a $2 M_\odot$ NS and the constraint on $\Lambda$ from GW170817 have been imposed. The trends of the slopes of EoSs in both scenarios have been found to be consistent with various RMF EoSs used in the literature~\cite{Xia:2022dvw}. The regions excluded by the condition $\mc{P} < 3(\rho^{V} m_{n}^{\star} + \rho^{V}_{\chi} m^{\star}_{\chi} + \mathcal{E}) = 3\mathcal{E}_{tot}$ from the relativistic kinetic theory considerations \cite{Olson:2000vx} fall always much above the EoS lines. The `matching' point on the EoS indicates the point where the crust EoS has been matched with the BHF-corrected EoS.

\par In Fig.~\ref{rx-my-pulsar-rmf} and Fig.~\ref{rx-my-pulsar-bhf}, we compare the mass-radius ($M$-$R$) relations for purely hadronic and DMANS scenarios, in both RMF and BHF cases respectively, with observational data from various pulsars listed in Table~\ref{tab:Data}. The $M$-$R$ relations are obtained by solving the TOV equations as shown in Eq.~\eqref{eq:tov} in Appendix~\ref{macro}. We find that, at a fixed NS mass, the BHF corrections significantly reduce the radius compared with the RMF results, thereby increasing the compactness of the star. The RMF model yields maximum masses and radii that are broadly consistent with relatively heavier pulsars and GW observations. In contrast, the BHF-corrected results show tighter mass-radius curves, and while they provide better agreement with low-mass compact stars such as PSR J0437-4715, PSR J1231-1411, and the ultracompact object HESS J1731-347, they are not in good agreement with the masses of heavier pulsars e.g. PSR J0740+6620. 
\par Similarly, in Fig.~\ref{rx-vs-my-gw}, we compare the same $M$-$R$ relations with various GW events. The BHF cases are not in good agreement with the mass estimates of GW190425 M1(M2)~\cite{LIGOScientific:2020aai}, whereas the RMF-based results offer a better match. However, for GW170817~\cite{LIGOScientific:2018cki} event, the BHF predictions remain consistent, albeit with smaller radii compared with the RMF case. We show two special regions, as pointed out in Ref.~\cite{Annala:2017llu}, of tidal deformability $\Lambda_{1.4}$ at $M_{\mathrm{NS}} = 1.4 M_\odot$ i.e. $\Lambda_{1.4} < 400$ and $400 < \Lambda_{1.4} < 800$. All the BHF curves lie within the region $\Lambda_{1.4} < 400$, which is consistent with the results shown later in Fig.~\ref{fig:M-Lambda}.

\begin{table*}
\centering
\renewcommand{\arraystretch}{1.0}  
\setlength{\tabcolsep}{7pt}       
\begin{tabular*}{\textwidth}{@{\extracolsep{\fill}} l c c c c c c}
\toprule\toprule
\multirow{2}{*}{Observables} & \multicolumn{3}{c}{RMF} & \multicolumn{3}{c}{BHF} \\ 
\cmidrule(lr){2-4} \cmidrule(lr){5-7} 
& Hadronic  & BPI & BPII & Hadronic  & BPI & BPII \\ 
\midrule\midrule
$M_{\mathrm{max}}$ ($M_\odot$) & $2.13$ & $2.08$ & $2.00$ & $2.03$ & $1.99$ & $1.93$ \\ 
$R_{\mathrm{peak}}$ (km) & $11.81$ & $11.34$ & $10.69$ & $10.41$ & $10.29$ & $10.04$ \\ 
$\Lambda_{1.4}$ & $532$ & $493$ & $336$ & $134$ & $125$ & $109$ \\ 
\bottomrule\bottomrule
\end{tabular*}
\caption{Maximum masses $M_{\mathrm{max}}$ of NSs, corresponding radii $R_{\mathrm{peak}}$, and tidal deformability at $M_{\mathrm{NS}} = 1.4 M_\odot$ (denoted by $\Lambda_{1.4}$) for purely hadronic and DMANS (BPI and BPII) in RMF and BHF frameworks.}
\label{tab:M-R-Lambda-Data}
\end{table*}

Interestingly, while Ref.~\cite{Sagun:2023rzp} required a large DM fraction (around $4.75$\%) to attain consistency with HESS J1731-347, our 
BHF-corrected DMANS model achieves good agreement with a significantly smaller DM fraction  of $\approx 0.2\%$ (or even no DM at all). The purely RMF-based hadronic model fails to fall within the radius estimate of PSR J1824-2452I (M28I), unlike its BHF counterpart. As the DM mass increases (for fixed $k_{\chi}$), both the maximum mass and corresponding radius of the NS decrease, consistent with previous studies~\cite{Xiang:2013xwa,Das:2021hnk,Lenzi:2022ypb}. However, we observed that for a fixed NS mass, the radius is less sensitive to $M_{\chi}$ in the BHF model compared with the RMF case. We note that hadronic configurations in the BHF scenario show better agreement with the transient low-mass X-ray binaries such as EXO 1745-248 and 4U 1608-52~\cite{Ozel:2015fia}, a feature not captured by the RMF-based hadronic model.

\par
Although within our BHF approach, more compact objects can be better explained compared with the RMF approach, the ultra--compact object XTE J1814-338~\cite{cbat2003} cannot be explained in our framework. In the context of RMF models, studies using bosonic dark matter models \cite{Pitz:2024xvh}, Higgs-mediated fermionic dark matter models \cite{Lopes:2024ixl}, mirror dark matter admixed strange star models \cite{Yang:2024ycl} and hybrid star models \cite{Laskos-Patkos:2024fdp} have all aimed to explain the ultracompact nature of XTE J1814-338. Given the lesser degree of sensitivity of stellar radius in the BHF case upon inclusion of DM, achieving consistency with XTE J1814-338 perhaps requires significantly larger DM masses compared with the RMF scenario.

\par In Fig.~\ref{fig:M-C}, we present the mass–compactness ($M$-$C$) relations for NSs within both the RMF and BHF frameworks, considering purely hadronic and DMANS. We find that the compactness is significantly less sensitive to variations in the DM mass in the BHF case compared with the RMF case. Using the derived compactness of PSR J0030+0451~\cite{Riley:2019yda}, we find that in the BHF scenario, the inferred mass range for this pulsar lies between $(1.03 - 1.19)M_\odot$, which is lower than the observationally estimated mass range for PSR J0030+0451. In contrast, the RMF-based result yields a more consistent mass range of $(1.19 - 1.49)M_\odot$. Conversely, if one uses the known mass estimate of PSR J0030+0451 to infer compactness, we obtain $C \in [0.13, 0.19]$ for the RMF case and $C \in [0.16, 0.20]$ for the BHF case. These results indicate that the BHF model is not fully compatible with both the mass and compactness constraints of PSR J0030+0451, while the RMF case provides a better fit. As for a particular BHF $M$-$R$ curve, the radius of the star is constrained in a very narrow range around $R = 10.9$ km, the compactness increases almost linearly with mass.

\par In Fig.~\ref{fig:M-Lambda}, we present the variation of the dimensionless tidal deformability $\Lambda$ as a function of NS mass for both RMF and BHF models, considering pure hadronic as well as DMANS scenarios. While the BHF results do not agree well with the benchmark value $\Lambda_{1.4} \approx 370^{+360}_{-130}$ inferred from the combined analysis of GW170817 and NICER observations of PSR J0030+0451~\cite{Jiang:2019rcw}, they remain well within the broader range $\Lambda_{1.4} \approx 70-580$ derived from the GW170817 data alone~\cite{LIGOScientific:2017vwq,LIGOScientific:2018cki}. We also compare theoretical results with $68\%$ confidence limits on GW190425 M1(M2)\cite{Traversi:2021fad}. At $M_{\mathrm{NS}} = 1.4M_\odot$, we observe that, for a fixed $k_\chi$, the tidal deformability $\Lambda$ in the BHF scenario exhibits reduced sensitivity to the DM mass compared with the RMF case. This reduced sensitivity is a consequence of the increased stiffness of the NS matter resulting from BHF corrections, as discussed earlier. For completeness, Table~\ref{tab:M-R-Lambda-Data} provides a summary of the key macroscopic observables--including the maximum mass $M_{\rm max}$, corresponding radius $R_{\rm peak}$ and $\Lambda_{1.4}$--for the benchmark DM parameter sets BPI and BPII, in both RMF and BHF frameworks.

\par The quantities $C$ and $\Lambda$ are connected via Eq.~\eqref{eq:TD}, where another quantity called tidal Love number $k_2$ is also present (see Appendix~\ref{macro} for its definition and related discussions). The variation of $k_2$ with DM mass in the RMF and BHF frameworks has been found to be qualitatively opposite. At a fixed stellar mass, $k_2$ mildly increases with DM mass in the RMF case, whereas it decreases mildly in the BHF case. Understanding the physical origin of this small variation of $k_2$ with respect to the DM mass is difficult, given its complex dependence on $C$ and the metric perturbation parameter $y_R$. However, such mild variations of $k_2$ also have marginal effects on $\Lambda$ compared with the effect of compactness $C$. The variation in $\Lambda$ is heavily dependent on the variation in $C$ as $\Lambda \sim C^{-5}$. In the RMF scenario, $C$ increases upon inclusion of DM (e.g. at $M = 1.4M_\odot$, $C$ is larger by $13\%$ for the BPII parameter set compared with the hadronic scenario). Given the strong dependence of $\Lambda$ on $C$, this leads to $\approx 45\%$ decrease in $\Lambda$ for the BPII set compared with the hadronic case. A similar trend is also observed within the BHF results. Additionally, for a fixed $M$, considerable increase of $C$ in the BHF case (by $\approx 17\%$ in the BHF hadronic case compared with RMF) leads to a large decrease in $\Lambda$ ($\approx$ 82$\%$) in the BHF case.

\par We present mass-MoI ($M$-$I$) relation in Figs.~\ref{fig:M_I(a)} in the slow-rotation approximation, where the rotation frequency of the star is always much smaller than the Kepler frequency and the MoI becomes largely independent of the rotation frequency of the star. Relevant expressions for the calculations of MoI are given in Appendix \ref{macro}. Although for a fixed $k_{\chi}$, the MoI decreases for increasing $M_{\chi}$ in both scenarios, variation of MoI is less sensitive to the DM mass in the BHF case. Similar behaviour has been observed w.r.t. radius.
Both the RMF and BHF cases are consistent with mass-MoI estimate of PSR J0030+0451~\cite{Jiang:2019rcw}. The mass-MoI estimate of light pulsar PSR J0737-3039A \cite{Landry:2018jyg,Miao:2021gmf} is more consistent with the BHF scenario. Fig.~\ref{fig:R_I(b)} presents radius-MoI (R-I) relations. The BHF case has found not to be in good agreement with radius-MoI estimate of PSR J00340+0451. The inferred MoI of PSR J0737-3039A constrains the radius in the BHF star radii smaller ($\approx 10.6-11$ kms) than the RMF star ($\approx 11.4-13.3$ kms). The BHF results are also consistent with the radius measurement of light pulsar PSR J1231-1411. We observe that for stars with identical mass, $I_{\rm BHF}$ turns out to be smaller than $I_{\rm RMF}$. For instance at $M = 1.4M_\odot$ (for pure hadronic matter), the ratios, $R_{\rm BHF}/R_{\rm RMF}\approx 0.8$, $\Omega_{\rm BHF}/\Omega_{\rm RMF}\approx 1.1$ and $\bar\omega^\prime_{\rm BHF}/\bar\omega^\prime_{\rm RMF}\approx 1.8$ (here $\bar\omega^\prime=d\bar\omega/dr$). These ratios give $I_{\rm BHF}/I_{\rm RMF}\approx 0.68$ which is consistent with the values obtained from Fig. \ref{fig:M_I(a)}.
\begin{figure*}
\centering
\captionsetup[subfigure]{labelformat=empty}
\subfloat[\quad(a)]{\includegraphics[height=9.0cm,width=9.0cm]{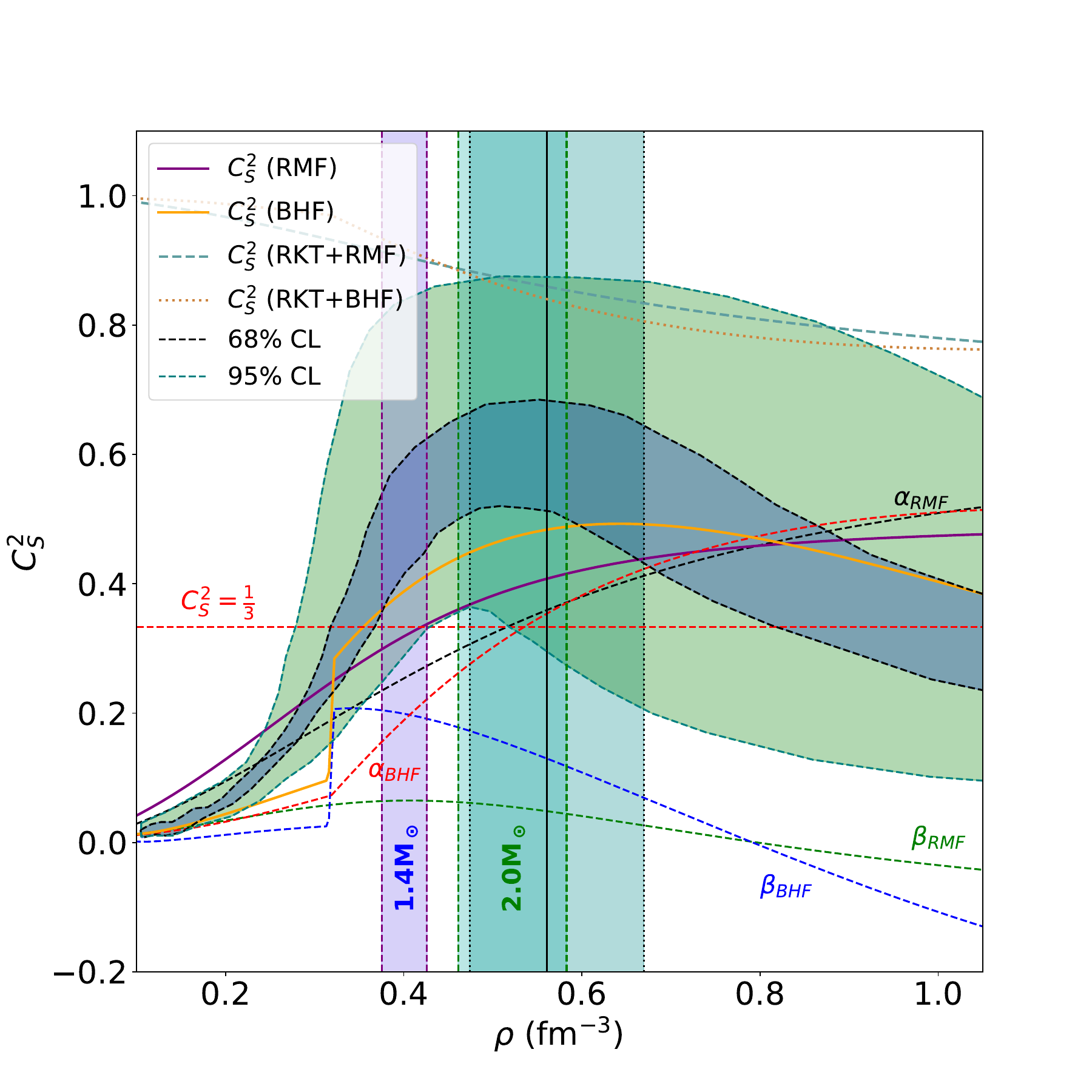}\label{fig:rho-cs2}}\hfill
\subfloat[\quad(b)]{\includegraphics[height=9.0cm,width=9.0cm]{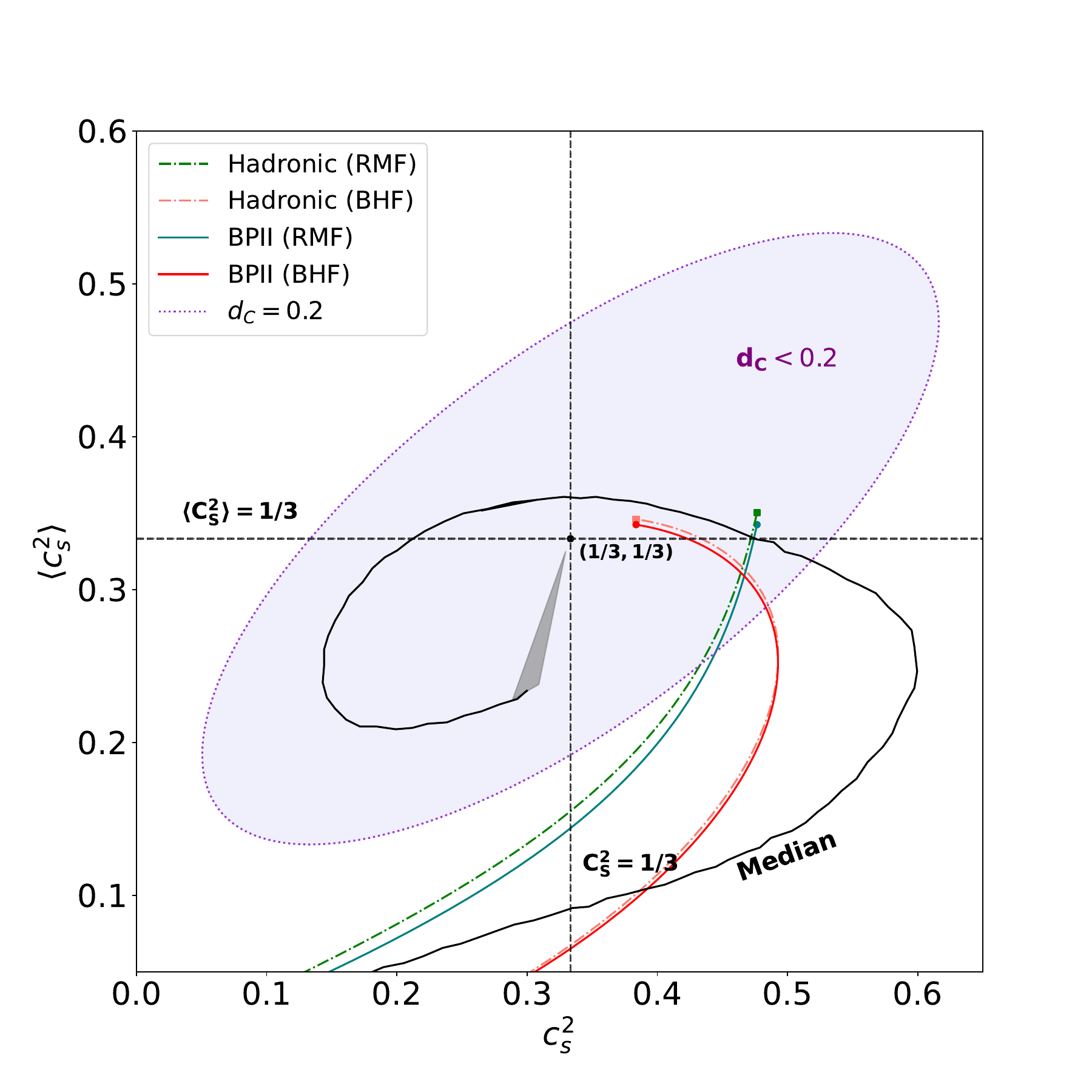}\label{fig:cs2_avcs2}}\\
\subfloat[\quad(c)]{\includegraphics[height=9.0cm,width=9.0cm]{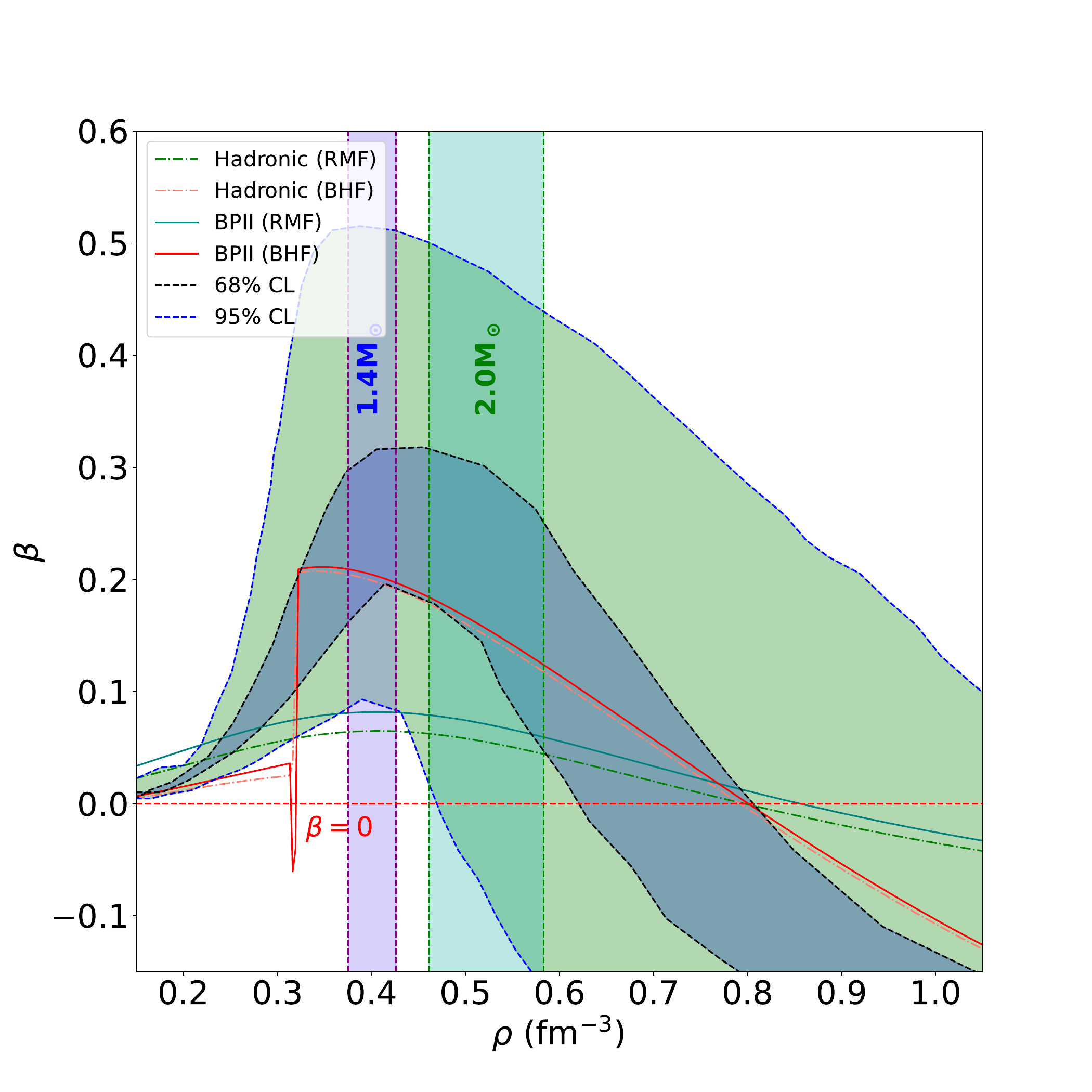}\label{fig:rho-beta}}\hfill
\subfloat[\quad(d)]{\includegraphics[height=9.0cm,width=9.0cm]{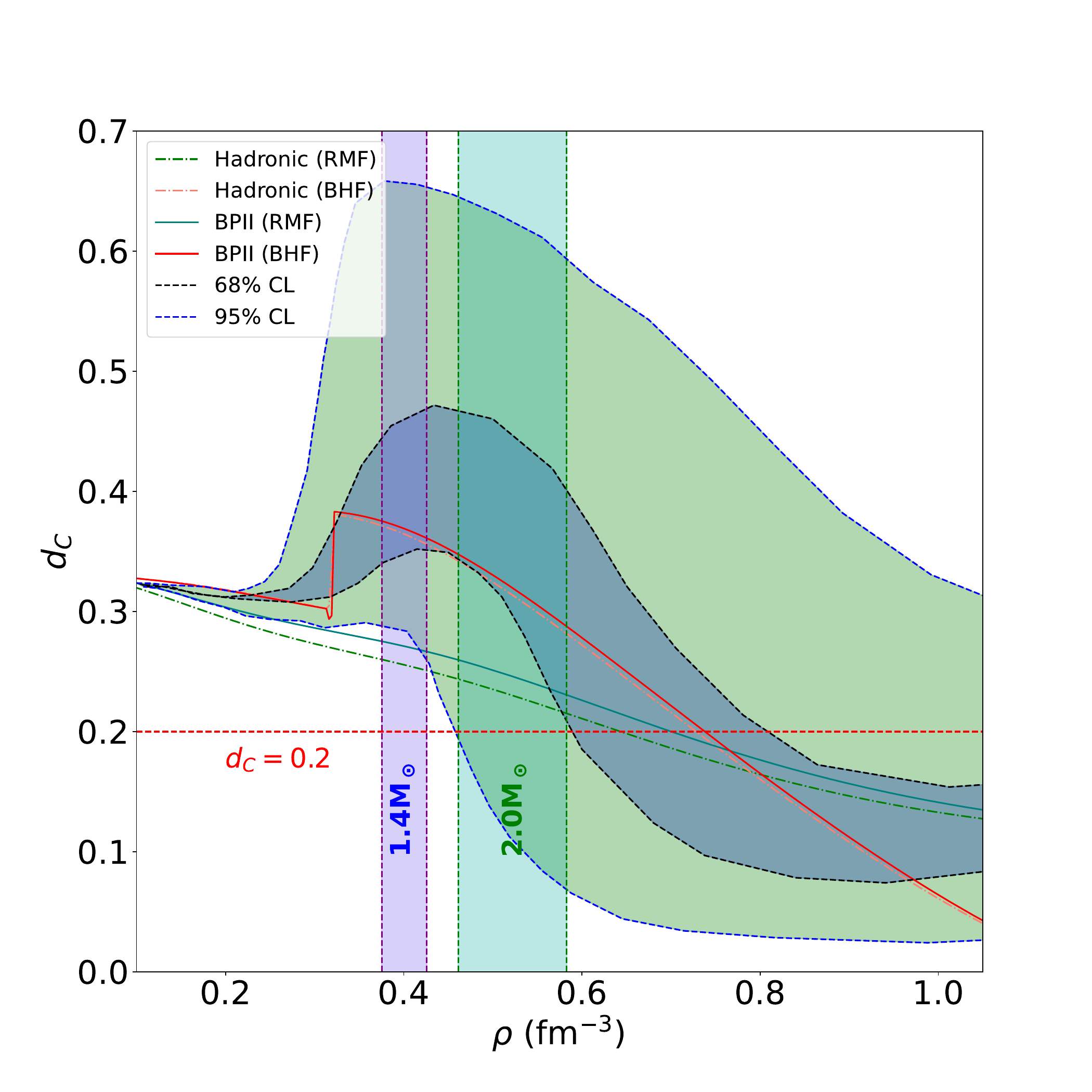}\label{fig:rho-dc}}
\caption{
(a) Comparison of $C^2_S$, $\alpha$ and $\beta$ \cite{Marczenko:2025hsh} in RMF and BHF cases as functions of $\rho^V$. The green vertical shaded region bounded by black dotted lines encloses the $1\sigma$ credible region for the peak of $C^2_S$ \cite{Marczenko:2022jhl}. The shaded regions bounded by purple and green vertical lines enclose $1\sigma$ credible intervals for central densities of $1.4\,M_\odot$ and $2.0\,M_\odot$ NSs \cite{Annala:2023cwx}. 
(b) EoSs on the $C^2_S - \langle C^2_S \rangle$ plane for hadronic and DMANS matter, with RMF and BHF curves. Colored dots mark $\rho^{V} \approx 1.05\,\mathrm{fm}^{-3}$ ($\approx 7\rho_0$). The solid black line shows the mean EoS from a statistical ensemble consistent with constraints in Ref.~\cite{Marczenko:2025hsh}. 
(c) Variation of $\beta$ as a function of $\rho^V$ for hadronic and DMANS cases. 
(d) Variation of conformal distance $d_c$ with \tcr{$\rho^V$}.
}
\label{fig:combined_cs2_beta_dc}
\end{figure*}

\section{Speed of Sound and Conformality}
\label{result:conformality}

The speed of sound ($C_S$) inside a NS is a direct measure of the stiffness of the EoS. This is defined as the slope of the EoS curve via
\begin{align}
\label{eq:sos}
C^2_S = \frac{\partial \mc{P}}{\partial \mc{E}}\,.
\end{align}
In the previous section, we observed that incorporating the BHF-corrected effective nucleon mass leads to significant differences in the $M$-$R$ relations, tidal deformability, compactness and MoI. 
Although $C_S$ cannot be directly measured, it can be inferred indirectly from EoS constraints and other NS observables obtained through pulsar and GW data. Bayesian methods are commonly used to infer $C_S$ from observational priors by sampling various EoS parametrisations. These studies suggest that to support the existence of $\approx 2M_\odot$ NSs, $C_S$ must exceed the conformal limit $C_S \leq c/\sqrt{3}$ (where $c$ is the speed of light in vacuum), typically peaking around 
$\rho \approx 3\rho_0-4\rho_0$, attaining values $C^2_S \approx 0.6$ \cite{Chatterjee:2023ecc}. These results strongly indicate possible stiffening of the EoS in the NS interior. We present a comparative analysis of the $C_S$ and associated measures of conformality in the RMF and BHF frameworks. We note that the inclusion of a small fraction of DM in the NS does not significantly alter $C_S$~\cite{Lopes:2024ixl}.

In Fig.~\ref{fig:rho-cs2}, we show the variation of $C^2_S$ as functions of $\rho^V$, along with the slope ($\alpha$) and curvature ($\beta$) of the energy per particle (see Appendix~\ref{conformality} for a brief discussion), in both the RMF and BHF frameworks. These theoretical predictions are compared with the confidence intervals derived in Ref.~\cite{Marczenko:2023txe}, in which a piecewise-linear speed-of-sound parametrisation was adopted for the EoS in the density range $\lt[0.5,1.1\rt]\rho_0$~\cite{Annala:2019puf}, while using the Baym-Pethick-Sutherland EoS at sub-nuclear densities and pQCD EoS at asymptotically high densities. The observational inputs used in their analysis include the tidal deformability $\Lambda_{1.4}$ from GW170817~\cite{LIGOScientific:2018cki} and the mass of PSR J0952–0607~\cite{Romani:2022jhd}, leading to constraints expressed as functions of energy density $\mc{E}$. For comparison with our results expressed in terms of $\rho^V$, we have rescaled the energy density values in their constraints by a factor of $\rho_0/\mc{E}_0$, where $\mc{E}_0 \approx 150~\text{MeV}/\text{fm}^3$ denotes the nuclear saturation energy density. 

The non-monotonic behaviour of $C^2_S$ in the BHF scenario is consistent with earlier studies~\cite{Kojo:2020krb,Altiparmak:2022bke,Ecker:2017fyh,Brandes:2022nxa}. At sufficiently high baryon densities, the breach of the conformal limit is also supported by existing literature~\cite{Altiparmak:2022bke,Chatterjee:2023ecc}. However, we note a tension between the radii of maximally massive stars ($M\approx 2M_\odot$) in the BHF case ($R = 10.4$ km in the purely hadronic case) and the corresponding values reported in Ref.~\cite{Altiparmak:2022bke}. In Ref.~\cite{Annala:2019puf}, it was shown that for neutron stars with mass $M\approx 2M_\odot$, if the conformal bound is not strongly violated, the stars are predicted to contain quark matter cores. In our BHF-based hadronic EoS, the maximum mass configuration occurs at central energy density $\mathcal{E}_c \approx 1412$ MeV/fm$^{3}$, corresponding to $\rho^V \approx 1.05$ fm$^{-3}$. This lies slightly outside the validity range of the BHF fitting function. Nonetheless, we extrapolate the BHF parametrisation up to this density and find that at high $\rho^V$, the BHF corrections bring $C^2_S$ close to the conformal limit. This behaviour could be indicative of a transition toward weakly coupled quark matter~\cite{Gorda:2018gpy,Gorda:2021znl}. Such a trend can be attributed to the large negative values of $\beta_{\rm BHF}$ at high $\rho^{V}$, as illustrated in Fig.~\ref{fig:rho-cs2}. Furthermore, the values of $C^2_{S}$ in the BHF framework lie within the $1\sigma$ confidence band presented in Ref.~\cite{Marczenko:2023txe} over the range $\rho^V \approx [0.6-1.05]$ fm$^{-3}$.

We also compare our results with the $1\sigma$ credible region for the location of the peak of $C^2_S$ obtained in Ref.~\cite{Marczenko:2022jhl}, as well as with the central densities corresponding to $1.4 M_\odot$ and $2.0 M_\odot$ NSs, as reported in Ref.~\cite{Annala:2023cwx}. To ensure that the $C_S$ in the BHF scenario remains causal, apart from $C^2_S < 1$, we also verify that it satisfies the density-dependent upper bound derived from relativistic kinetic theory~\cite{Olson:2000vx}, (shown by RKT + RMF/BHF in the plot where "RKT" is used as an abbreviation for "\textbf{r}elativistic \textbf{k}inetic \textbf{t}heory"). The constraint is given by
\begin{align}
C_S^2 \leq \frac{\mc{E} + \rho^V m_n^{\star} + \rho_\chi^{V} m_{\chi}^{\star} - \mc{P}/3}{\mc{E} + \rho^V m_n^{\star} + \rho_\chi^{V} m_{\chi}^{\star} + \mc{P}}
\label{CS2Kin}
\end{align}
We have used the appropriate density-dependent effective masses for all fermionic species to evaluate this inequality.

\par From Eq.~\eqref{eq:sos} and Eq.~\eqref{avcs2}, the EoSs can be obtained on the $C^2_{S}-\langle C^2_{S} \rangle$ plane as shown in Fig.~\ref{fig:cs2_avcs2} in both RMF and BHF scenarios for both hadronic and DMANS matter. The contour $d_c = 0.2$ \cite{Marczenko:2025hsh} (refer to Appendix \ref{conformality} for a brief discussion on various measures of conformality) separates non-conformal ($d_c > 0.2$) and nearly conformal ($d_c \leq 0.2)$ behaviour. 
Near $\rho^{V} \approx 7\rho_0 \approx 1.05$ fm$^{-3}$, the BHF case is much closer to full conformality (the point of intersection of the lines $C^2_S = 1/3$ and $\langle C^2_S \rangle = 1/3$) than the RMF case. Similarly, one can get the EoS on the $\Delta-C^2_S$ plane to arrive at the same conclusion. 
Moreover, unlike the RMF scenario, the BHF EoS in the latter case has been found to be consistent with the 1$\sigma$ interval of $\Delta$ as a function of $C^2_S$ at the center of maximally stable heavy NSs~\cite{Marczenko:2022jhl}. 

\par In Fig.~\ref{fig:rho-beta}, we show the variation of $\beta$ as functions of $\rho^V$. Conformality sets in for both RMF and BHF around the same density of $\rho^V \approx 0.8$ fm$^{-3}$. At even higher densities, both are consistent with the $2\sigma$ confidence interval. Moreover, unlike RMF, within the intermediate density range of $[0.3,0.6]$ fm$^{-3}$, the BHF scenario is consistent with the $1\sigma$ confidence interval.
An enhanced values of $\beta$ in the BHF case in the range $\rho^V \in [0.3-0.8]$ fm$^{-3}$ implies an enhanced value of compression modulus \cite{Ivanytskyi:2022bjc}.
\par In Fig.~\ref{fig:rho-dc}, we show the variation of the conformal distance $d_c$~\cite{Annala:2023cwx} as functions of $\rho^V$.~In the density range of $[0.3-0.95]$~fm$^{-3}$, the BHF scenario is more consistent with the $1\sigma$ limits than RMF. However, for densities beyond $0.95$~fm$^{-3}$, this trend is reversed.

\par Besides these results, we have also studied the variation of the polytropic index ($\gamma$) with $\rho^V$. In the range of $[0.4-0.7]$~fm$^{-3}$, the BHF case has been found to be much more consistent with the $1\sigma$ limit~\cite{Marczenko:2023txe} than the RMF case. At even higher densities, both cases are consistent with $2\sigma$ confidence limit. For the BHF case, the onset of conformality occurs ($\gamma \approx 1.75$, a necessary but not a sufficient condition \cite{Somasundaram:2021clp}) at a baryon density slightly higher than the RMF case. We have also studied the variation of the conformal distance $d_c$ with $C^2_S$. The RMF EoS crosses the $d_c = 0.2$ conformality threshold at lower values of $C^2_S$ compared with the BHF case. In the high-density limit, BHF cases are much closer to full conformality than the RMF cases, but at a lower $C^2_S$ compared with the RMF.

\section{Conclusion and Discussion}
\label{END}

In this work, we have studied the properties of ultradense matter, speed of sound and various aspects of conformality in NS with and without an admixture of DM, using a relativistic mean-field model improved by the inclusion of realistic medium effects of many-body interactions.
Specifically, we implemented BHF-informed effective nucleon masses within the RMF framework~\cite{Shang:2020kfc} to model interacting fermionic DM admixed with normal nuclear matter described by the $\mathrm{SU(2})$ chiral sigma model~\cite{Guha:2021njn}. We have shown the effects of such many-body interactions on macroscopic observables, speed of sound and conformality of purely hadronic and DMANS. 

\par The BHF-corrected EoS for static neutron stars lowers the maximum mass ($1.9M_\odot-2.0M_\odot$) compared with the purely RMF case ($2M_\odot-2.2M_\odot$). The inclusion of BHF corrections increases the compactness of the star, allowing for improved consistency with smaller compact objects e.g. HESS J1731-347 and PSR J1231-1411. In contrast with Ref.~\cite{Sagun:2023rzp}, our results are in agreement with HESS J1731-347 with much smaller density of DM. In the BHF scenario, as the DM mass increases, the maximum masses and the corresponding radii decrease; also the radii are less sensitive to DM mass compared with the RMF case. Large DM masses (e.g. 15 GeV or more) lead to tension with the mass of very heavy pulsars but is still consistent with relatively low mass compact objects e.g. HESS J1731-347. Dimensionless tidal deformabilities ($\Lambda$) are consistent with measurements on GW170817~\cite{LIGOScientific:2017vwq,LIGOScientific:2018cki} but in tension with joint analysis of PSR J0030+0451 and GW170817~\cite{Jiang:2019rcw}. These effects demonstrate that current DM admixed RMF models which do not include many-body effects between nucleons need to be improved if they are to achieve consistency with data on heavy NSs. The $M$-$R$ relations obtained within the BHF framework are well within the constraint $\Lambda(1.4 M_\odot) < 400$ \cite{Annala:2017llu}, a result confirmed by the corresponding $M$-$\Lambda$ relations.

\par The BHF correction also changes the behaviour of the speed of sound by making its profile non-monotonic~\cite{Kojo:2020krb,Altiparmak:2022bke,Ecker:2017fyh,Brandes:2022nxa} and improving agreement in the high density regime with the available confidence limits from Ref.~\cite{Marczenko:2023txe}. Such behaviour of $C^2_S$ could be indicative of the onset of weakly coupled quark matter phase~\cite{Gorda:2018gpy,Gorda:2021znl}. For both RMF and BHF scenarios, the $C^2_S$ profile is consistent with constraints from relativistic kinetic theory \cite{Olson:2000vx}.
The BHF-corrected EoS approaches conformality more closely than the RMF case near \( \rho^V_n \approx 7\rho_0 \approx 1.05 \) fm$^{-3}$, as illustrated on the $C_S^2$-$\langle C^2_S \rangle$ plane. BHF results better match the \( 1\sigma \) confidence limits for conformality parameters e.g. \( \beta \)~\cite{Annala:2023cwx} and \( d_c \)~\cite{Marczenko:2023txe}.

\par A brief discussion of some possible extensions of the present work are in order. The construction of the $\mathrm{SU(2)}$ chiral sigma model used in this work employs nucleons that are degenerate in mass. A more realistic model with isospin-breaking effect by introducing an isovector scalar field could be considered .  
Different effective-mass profiles for $n$ and $p$ can be used \cite{Shang:2020kfc} in this scenario to obtain more refined results. Effects of couplings between mesons, rapid rotation and finite temperatures would further influence the present results, as would the introduction of conjectured additional degrees of freedom such as hyperons along with nucleon-hyperon three-body interactions. Effects of external parameters such as \tcr{temperature and} strong magnetic field on $\beta$-equilibrated, charge neutral nuclear and quark matter~\cite{Mandal:2009uk,Mandal:2012fq,Mandal:2016dzg,Mandal:2017ihr} in the context of NS observables might be an interesting avenue to investigate in future. Moreover, because of differences in the speed of sound profile of the BHF scenario compared with the RMF, it would be interesting to investigate the effects of such differences in the non--radial modes of oscillations of NS, such as $r$--modes~\cite{Jaikumar:2008kh} and $g$--modes~\cite{Jaikumar:2021jbw}. 

\section*{acknowledgement}
T.M. acknowledges partial support from the SERB/ANRF,
Government of India, through the Core Research Grant
(CRG) No. CRG/2023/007031. P.J. is supported by a grant from the U.S. National Science Foundation PHY-2310003. A.D. thanks Dr. Sanchari Goswami for carefully reading the manuscripts and sharing her valuable comments on the paper.
\appendix
\section{EoS and EoM}
\label{app:eos_eom}

\noindent
The quantity $\Sigma$ introduced in Eq.~\eqref{eq:PE} is defined as follows:
\begin{align}
\label{Cap_sigma}
\Sigma = \dfrac{g_{\sigma}^2x_{0}^2}{C_\sigma}\left[\dfrac{\overline{Y}^4}{8}-\dfrac{B\overline{Y}^6}{12C_\omega}+\dfrac{C\overline{Y}^8}{16C_\omega^2}\right],
\end{align}
where $\overline{Y}^2=1-Y^2$. This captures the contribution to the EoS coming from the scalar potential $U(\bar{x})$. Using the relations, $C_{\sigma} = g^2_{\sigma}/m^2_{\sigma}$ and $C_{\omega} = g^2_{\omega}/m^2_{\omega} = x_0^{-2}=f_\pi^{-2}$, the above expression for $\Sigma$ simplifies to,
\begin{align}
\label{sigma_simp}
\Sigma = m^{2}_{\sigma}f^{2}_{\pi}\left[\frac{\overline{Y}^4}{8} - \frac{f^2_{\pi}B\overline{Y}^6}{12} + \frac{f^4_{\pi}C\overline{Y}^8}{16}\right].
\end{align} 
In the above, we have used the nucleon effective mass ratio $Y$ defined as  
\begin{align}
Y = \dfrac{g_{\sigma}\sigma + g_{\phi}\phi}{m_N}.
\end{align}
The effective nucleon mass is given by, $m_N^{\ast} = Ym_N$. Similarly, the effective DM mass ratio is given by
\begin{align}
Y_{DM} = \frac{m_{\chi} + y_{\phi}\phi}{m_{\chi}}.
\end{align}

The quantity $\Theta_{\pm}$,  introduced in Eq.~\eqref{eq:PE} is defined as follows:
\begin{align}
\label{Cap_Theta}
\Theta_{\pm} = \frac{C_{\omega}{\lt(\rho^V\rt)}^2}{2Y^2} + \frac{1}{2}m^2_{\rho}\lt(\rho^{03}\rt)^2 + \frac{1}{2}m^2_{\xi}\lt(\xi^{0}\rt)^2 \pm \frac{1}{2}m^2_{\phi}\phi^2,
\end{align}
where the first (second) index on $\rho$ denotes the Lorentz [$\mathrm{SU}(2)$] index. All other components of $\vec{\rho}_{\mu}$ are zero. This captures contributions coming from the mass terms of various mesons. In this expression $\rho^V = \rho^{V}_{n} + \rho^{V}_{p}$. However, in the case of a hadronic sector composed solely of neutrons, $\rho^V = \rho^{V}_{n}$ (since $\rho^{V}_{p}=0$).
The EoM for $\phi$ is
\begin{equation}
    \phi = - \frac{g_{\phi}}{m^2_{\phi}} \rho^{S} - \frac{y_{\phi}}{m^2_{\phi}} \rho^{S}_{\chi},
    \label{eom:phi}
\end{equation}
where $\rho^S = \rho^S_n + \rho^S_p$. Using the familiar results of $\langle \overline{\psi} \gamma^{i} \psi \rangle = 0 = \langle \overline{\chi} \gamma^{i} \chi \rangle$ ($i = 1,2,3$) at zero temperature \cite{Glendenning:1997wn}, and hence, $\xi^{i} = 0$, we get the EoM for $\xi^{0}$ as
\begin{align}
    \xi^{0} &= \frac{g_{\xi}}{m^2_{\xi}}\rho^V +\frac{y_{\xi}}{m^2_{\xi}}\rho^{V}_{\chi}\nn\\ &
    = \frac{g_{\xi}}{m^2_{\xi}}
    \left(\frac{\gamma_{n}}{6\pi^2}k_{n}^3 + \frac{\gamma_{p}}{6\pi^2}k_{p}^3\right) +\frac{y_{\xi}}{m^2_{\xi}}\frac{\gamma_{\chi}}{6\pi^2}k^{3}_{\chi}.
\end{align}
Here, $\rho^V = \rho^V_n + \rho^V_p$. Similar reasoning leads to $\omega^{i} = 0$ and we get the EoM of $\omega^{0}$ as
\begin{align}
    \omega^{0}  = \frac{\rho^V}{g_{\omega}\sigma^2} = \frac{1}{g_\omega \sigma^2}\bigg(\frac{\gamma_{p}}{6\pi^2}k_{p}^3 + \frac{\gamma_{n}}{6\pi^2}k_{n}^3\bigg).
\end{align}
Similar assumptions lead to 
\begin{align}
    \rho^{03} = \frac{g_{\rho}}{2m^{2}_{\rho}}(\rho^V_n - \rho^V_p) = \frac{g_{\rho}}{2m^{2}_{\rho}}\bigg(\frac{\gamma_{n}}{6\pi^2}k_{n}^3 - \frac{\gamma_{p}}{6\pi^2}k_{p}^3 \bigg).
\end{align} 
The EoM for $\sigma$ is as follows:
\begin{align}
&~ \frac{2C_{\sigma}\rho^S}{Y(m_N - g_{\phi}\phi_{0})} 
- \frac{2C_{\sigma}C_{\omega}{\rho^V}^2}{Y^4(m_N - g_{\phi}\phi_{0})^2} -\overline{Y}^2 + \frac{B}{C_{\omega}}\overline{Y}^4 - \frac{C}{C^2_{\omega}}\overline{Y}^6 = 0.
\label{eom:sigma}
\end{align}
Using the expression for $\rho^{V}_{n}$ form 
Eq.~\eqref{rho_v} and using the EoMs for $\rho^{03}$ and $\xi^{0}$ (shown above) in the pure neutron matter case, the expression of $\Theta_{\pm}$ simplifies to
\begin{align}
\label{theta_simp}
    \Theta_{\pm} = \frac{\gamma^2_{N}C_{\omega}k^{6}_{n}}{72\pi^4Y^2} +  \frac{\gamma^{2}_{N}C_{\rho}k^{6}_{n}}{288\pi^4} + \frac{(\gamma_{N}g_{\xi}k^{3}_{n} + \gamma_{\chi}y_{\xi}k^{3}_{\chi})^2}{72\pi^4m^{2}_{\xi}} \pm \frac{1}{2}m^{2}_{\phi}\phi^2.
\end{align}

\section{Macroscopic properties of Neutron Stars}
\label{macro}

The mass-radius ($M$-$R$) relation of neutron stars is obtained by numerically solving the TOV equations, expressed here in gravitational units:
\begin{align}
\label{eq:tov}
\frac{d\mc{P}}{dr} &= -\frac{(\mc{E} + \mc{P})(\mc{M} + 4\pi r^3 \mc{P})}{r(r - 2\mc{M})}, \nn \\
\frac{d\mc{M}}{dr} &= 4\pi r^2 \mc{E},
\end{align}
where $\mc{M}(r)$ is the enclosed mass within radius $r$, $\mc{P}(r)$ is the pressure, and $\mc{E}(r)$ is the energy density. These equations are integrated simultaneously with the following boundary conditions: at the center of the star ($r = 0$), the pressure and mass satisfy $\mc{P}(0) = \mc{P}_c$ and $\mc{M}(0) = 0$, where $\mc{P}_c$ is the central pressure (or equivalently, $\mc{E}_c$ is the central energy density). The surface of the star is defined by the condition $\mc{P}(R) = 0$, which determines the stellar radius $R$. The stellar mass is obtained from $\mc{M}(r=R)=M$. 

The compactness of a neutron star (NS) is a dimensionless quantity, defined as
\begin{align}
\label{eq:C}
C = \frac{GM}{c^2R},
\end{align}
where $G$ is Newton’s gravitational constant, $c$ is the speed of light in vacuum, $M$ is the mass of the star, and $R$ is its radius. Since this is a dimensionless quantity, its numerical value remains the same regardless of the unit system. To obtain the compactness at an arbitrary radius $r$ within the star, one replaces $M\to\mc{M}(r)$ and $R\to r$ in the above expression.

To calculate the MoI of a NS, one must take into account its rotation. In a rotating NS, the centrifugal force experienced by a fluid element depends not only on the angular velocity $\Omega$ of the star, but also on the local frame-dragging angular velocity $\omega(r)$ at a radial coordinate $r$. The relative angular velocity is defined as:
\begin{align}
\bar{\omega}(r) = \Omega - \omega(r).
\end{align}
In the slow-rotation approximation, $\bar\omega(r)$ satisfies the following second-order ordinary differential equation (ODE)~\cite{Hartle:1967he}:
\begin{align}
\label{eq:omegabar}
\frac{1}{r^4} \frac{d}{dr} \left( r^4 j(r) \frac{d\bar{\omega}}{dr} \right) + \frac{4}{r} \frac{dj(r)}{dr} \bar{\omega} = 0,
\end{align}
where $j(r) = e^{-\nu}\sqrt{1 - 2\mc{M}(r)/r}$ for $r < R$ and $j(r) = 1$ for $r\geq R$. Also, $\nu(r)$ satisfies the following ODE
\begin{align}
\label{nu}
\frac{d\nu}{dr} = \frac{\mc{M}(r) + 4\pi r^3 \mc{P}(r)}{r \left[r - 2\mc{M}(r)\right]}.
\end{align}
The rotation frequency $\Omega(R)$ at the surface of the star is related to $\bar\omega(r)$ by the matching condition:
\begin{align}
\Omega(R) = \bar{\omega}(R) + \frac{R}{3} \left( \frac{d\bar{\omega}}{dr} \right)_{r=R}.
\end{align}
To solve Eq.~\eqref{eq:omegabar}, one imposes regular boundary conditions: $\left(d\bar\omega/dr\right)_{r=0} = 0$ and $\bar{\omega}(0)=\bar{\omega}_c$, 
where $\bar{\omega}_c$ is an arbitrarily chosen central value. Since Eq.~\eqref{eq:omegabar} is linear in $\bar{\omega}$, the solution can be scaled  to match any desired $\Omega(R)$. The MoI can be calculated in the slow rotation approximation using the following expression~\cite{Glendenning:1997wn}:
\begin{align}
\label{eq:moi}
I = \frac{c^2}{6G} \left[ \frac{R^4}{\Omega(R)} \left( \frac{d\bar{\omega}}{dr} \right)_{r=R} \right].
\end{align}
Although there is arbitrariness in the choice of the initial value $\bar{\omega}_c$, this does not affect the final value of $I$ in the slow rotation approximation, since both 
$\Omega(R)$ and $(d\bar\omega/dr)_{r=R}$ scale linearly with $\bar{\omega}_c$, and their ratio remains invariant.

\par Starting from the following unperturbed spherically symmetric, static metric 
\begin{eqnarray} \label{eq:rotmet}
   {ds}^{2} = - e^{\nu(r)} {dt}^{2}  + e^{\lambda(r)} {dr}^{2}  +  {r}^{2}({d\theta}^{2}  +  {sin}^{2} \theta {d\phi}^{2}),
\end{eqnarray}
one can perform a perturbative analysis on it to calculate $\Lambda$. The perturbed metric component $y(r)$ satisfies the following first order nonlinear ODE
\begin{equation}
    r\frac{dy}{dr} + y^{2} + y e^{\lambda} [ 1 + 4\pi r^{2} (\mc{P} - \mc{E})] + r^{2}Q = 0,
    \label{eq:pert}
\end{equation}
where 
\begin{equation}
    Q(r) = 4 \pi e^{\lambda} \: \bigg(5\mc{E}  + 9\mc{P} + \frac{\mc{E} + \mc{P}}{C_S^2}\bigg) - \frac{6e^{\lambda}}{r^{2}} - \bigg(\frac{d\nu}{dr}\bigg)^{2},
    \label{eq:Q}
\end{equation}
Using the boundary condition $y(0) = 2$, we calculate the tidal Love number $k_2$ ~\cite{Hinderer:2007mb,Hinderer:2009ca} as 
\begin{align}
k_2 &= \frac{8}{5} C^{5} ( 1- 2C)^{2} [ 2- y_{R} + 2C (y_{R} - 1)] 
 \times [2C\{6-3y_{R} + 3C \notag \\
    &\times (5y_{R} - 8)\}
 + 4 C^{3} \{ 13 - 11y_{R} + C (3y_{R} -2) + 2 
 C^{2}(1 + y_{R})\} \notag \\
    &+ 3( 1 - 2C)^{2}\{2 - y_{R} + 2C(y_{R} - 1)\} \ln(1-2C)]^{-1}.
    \label{eq:Love}
\end{align}
Here, $y_{R} = y(r=R)$ is the value of the perturbed metric element at the surface of the star and $\Lambda$ is given by~\cite{Leung:2022wcf}:
\begin{equation}
   \Lambda = \frac{2}{3}k_2 C^{-5} = \frac{2}{3}k_2\bigg(\frac{Rc^2}{GM}\bigg)^5. 
   \label{eq:TD}
\end{equation}
\par For the static NS, one solves the set of equations~\eqref{eq:tov}, \eqref{nu} and \eqref{eq:pert}. However, in the slowly rotating NS case, one solves Eqs.~\eqref{eq:tov},~\eqref{eq:omegabar}, \eqref{nu} and \eqref{eq:pert} simultaneously to calculate all the relevant observables.

\section{Measures of conformality}
\label{conformality}
Classical massless QCD turns out to exhibit conformal symmetry. Such a symmetry is characterized by the matter part of the trace anomaly~\cite{Fujimoto:2022ohj} $\langle \Theta \rangle = \langle \mc{T}^{\mu}_{\mu}\rangle = \mc{E} - 3\mc{P} = 0$ where $\mc{T}$ is the stress energy tensor of the model. So, fully conformal matter is characterized by the EoS $\mc{P} = \mc{E}/3$. However, quantizing the theory breaks this symmetry at the loop level due to finite masses of quarks. This deviation from conformality is measured by the deviation of $\langle \Theta \rangle$ from zero. As proposed in Ref.~\cite{Fujimoto:2022ohj}, however, we use 
\begin{eqnarray}
    \Delta = \frac{1}{3} - \frac{\mc{P}}{\mc{E}} = \frac{\langle \Theta \rangle}{3\mc{E}},
    \label{eq:delta}
\end{eqnarray}
as a measure of trace anomaly. Thermodynamic stability ($\mc{P} \geq 0$) and causality ($0 \leq C^2_S < 1$) requires that $-2/3 < \Delta < 1/3$. The average $\langle C^2_{S} \rangle$, in an energy density interval $[0,\mc{E}]$ is defined as
\begin{eqnarray}
    \langle C^2_{S} \rangle = \frac{1}{\mc{E}}\int^{\mc{E}}_{0}d\mc{E}^{\prime}\frac{d\mc{P}}{d\mc{E}^{\prime}} = \frac{\mc{P}(\mc{E})}{\mc{E}}.
    \label{avcs2}
\end{eqnarray}
This assumes a uniform probability distribution of $\mc{E}$. This leads to an alternate expression of $\langle C^2_S \rangle$ as follows $\langle C^2_{S} \rangle = 1/3 - \Delta$.
There are two different representations of $C^2_S$ that one can use in the present context. One can decompose $C^{2}_{S}$ as a sum of slope ($\alpha$) and curvature ($\beta$) of energy per particle $\mc{E}/\rho$ as follows~\cite{Marczenko:2023txe}
\begin{equation}
    C^{2}_{S} = \frac{1}{\mu}\frac{d\mc{P}}{d\rho} = \frac{2\rho}{\mu}\frac{d(\mc{E}/\rho)}{d\rho} + \frac{\rho^2}{\mu}\frac{d^2(\mc{E}/\rho)}{d  \rho^2} = \alpha + \beta,
\end{equation}
where $\mu$ is the baryon chemical potential. The quantities $\alpha$ and $\beta$ are defined as
\begin{equation}
    \alpha = \frac{2\rho}{\mu}\frac{d(\mc{E}/\rho)}{d\rho} =  \frac{2C^2_{S}}{C^2_{S} + \gamma} = 2\frac{1/3 - \Delta}{4/3 - \Delta},
\end{equation}
and
\begin{equation}
    \beta = \frac{\rho^2}{\mu}\frac{d^2(\mc{E}/\rho)}{d\rho^2} = C^2_{S} - \alpha.
\end{equation}
 The polytropic index $\gamma$ is defined as
\begin{equation}
    \gamma = \frac{\mc{E}}{\mc{P}}C^2_{S} = \frac{C^2_{S}}{\langle C^2_{S} \rangle} = \frac{C^2_{S}}{1/3 - \Delta}.
\end{equation}\\
An alternate representation of $C^2_{S}$ is as follows
\begin{equation}
C^2_{S} = \frac{1}{3} - \Delta - \Delta^{\prime},
\end{equation}
where
\begin{equation}
    \Delta^{\prime} = \mc{E}\frac{d\Delta}{d\mc{E}} = \langle C^2_{S} \rangle - C^2_{S}.
\end{equation}
The quantities $\Delta$ and $\Delta^{\prime}$ are called non-derivative and derivative contributions to the speed of sound, respectively.
The conditions $-2/3 < \Delta < 1/3$ and $C^2_S \in [0,1]$ lead to $\alpha\in [0,1]$ and $\beta \in [-1,1]$. At very low values of $\rho^V$, conformal symmetry is broken, i.e. $C^2_S \rightarrow 0$ and $\Delta \rightarrow 1/3$. This implies that $\alpha \rightarrow 0$ and $\beta \rightarrow C^2_S$. On the other hand, at asymptotically high $\rho^V$, conformal symmetry is restored, i.e. $\Delta \rightarrow 0$, and hence, $C^2_{S} \rightarrow 1/3$ and $\langle C^2_{S} \rangle \rightarrow 1/3$. This leads to $\alpha \rightarrow 1/2$ and $\beta \rightarrow -1/6$. Hence, in the asymptotically large baryon density, $\Delta \rightarrow 0$ and $\Delta^{\prime} \rightarrow 0$ in the fully conformal limit. This implies that the conformal distance~\cite{Annala:2023cwx},
\begin{equation}
d_c = \sqrt{\Delta^2 + \Delta^{\prime^2}} = \sqrt{\left(1/3 - \langle C^2_S \rangle\right)^2 + \left( C^2_S - \langle C^2_S \rangle\right)^2 },
\end{equation}
vanishes in the fully conformal limit. However, it is useful to ask for values of various parameters described above (and the corresponding baryon density around which such values are reached), which can characterize the onset of conformality. This could prove to be useful as intermediate densities where conformality sets in could be close to densities reached in NSs. It has been shown in the literature that $\Delta = 0$~\cite{Fujimoto:2022ohj}, $\beta = 0$~\cite{Marczenko:2023txe}, $\gamma \approx 1.75$~\cite{Annala:2019puf}, and $d_c \approx 0.2$~\cite{Annala:2023cwx} serve as fairly good threshold conditions for the onset of conformality.

\def\bibfont{\small}
\bibliography{First_Draft_Ref}{}

\end{document}